\documentclass[12pt, draftclsnofoot, onecolumn]{IEEEtran}
\makeatletter
\def\ps@headings{%
\def\@oddhead{\mbox{}\scriptsize\rightmark \hfil \thepage}%
\def\@evenhead{\scriptsize\thepage \hfil \leftmark\mbox{}}%
\def\@oddfoot{}%
\def\@evenfoot{}}
\makeatother 
\usepackage{amsfonts,dsfont}
\usepackage[dvips]{graphicx}
\usepackage{caption}
\usepackage{subfigure}
\usepackage{times}
\usepackage{cite}
\usepackage{lettrine}
\usepackage{amsmath}
\usepackage{bm}
\allowdisplaybreaks[4]
\usepackage{array}
\usepackage{amssymb}

\usepackage{stfloats}
\usepackage{slashbox}
\usepackage{graphicx}
\usepackage{footnote}
\usepackage{booktabs}
\usepackage{array}
\usepackage{subeqnarray}
\usepackage{cases}
\usepackage{threeparttable}
\usepackage{color}
\makeatletter
\renewcommand{\fnum@figure}{Fig. \thefigure}
\makeatother

\makeatletter
\newif\if@restonecol
\makeatother

\usepackage[linesnumbered,ruled,vlined]{algorithm2e}
\usepackage{algpseudocode}
\newcommand{\mc}[1]{\mathcal{#1}}
\newcommand{\bs}[1]{\boldsymbol{#1}}

\makeatletter
\renewcommand{\fnum@figure}{Fig. \thefigure}
\makeatother

\include{header}

\newtheorem{theorem}{{Theorem}}

\newtheorem{proposition}{{Proposition}}

\newtheorem{definition}{Definition}
\newtheorem{example}{Example}
\newtheorem{proof}{Proof}[section]

\newcommand{\tabincell}[2]{\begin{tabular}{@{}#1@{}}#2\end{tabular}}
\def\BibTeX{{\rm B\kern-.05em{\sc i\kern-.025em b}\kern-.08em T\kern-.1667em\lower.7ex\hbox{E}\kern-.125emX}}

\begin{document}
\title{Trajectory Design for Multiple-UAV Assisted \\ Wireless Networks}

\author{Yao Tang, Man Hon Cheung, and Tat-Ming Lok
\thanks{Part of this paper was presented in \cite{MyGC}. Yao Tang, Man Hon Cheung, and Tat M. Lok are with the Department of Information
Engineering, the Chinese University of Hong Kong (CUHK), Shatin, N.T., Hong
Kong.
E-mail: \{ty018, mhcheung, tmlok\}@ie.cuhk.edu.hk.
}
}

\maketitle


\thispagestyle{empty} 

\begin{abstract}
  Unmanned aerial vehicles (UAVs) can enhance the performance of cellular networks, due to their high mobility and efficient deployment.
  In this paper, we present a first study on how the \emph{user mobility} affects the UAVs' trajectories of a multiple-UAV assisted wireless communication system.
  Specifically, we consider the UAVs are deployed as aerial base stations to serve ground users who move between different regions.
  We maximize the throughput of ground users in the downlink communication by optimizing the UAVs' trajectories, while taking into account the impact of the user mobility, propulsion energy consumption, and UAVs' mutual interference.
  We formulate the problem as a route selection problem in an acyclic directed graph. Each vertex represents a task associated with a reward on the average user throughput in a region-time point, while each edge is associated with a cost on the energy propulsion consumption during flying and hovering.
  For the \emph{centralized} trajectory design, we first propose the shortest path scheme that determines the optimal trajectory for the single UAV case.
  We also propose the centralized route selection (CRS) scheme to systematically compute the optimal trajectories for the more general multiple-UAV case.
  Due to the NP-hardness of the centralized problem, we consider the \emph{distributed} trajectory design that each UAV selects its trajectory autonomously.
  We formulate the UAVs' interactions as a route selection game. We prove that it is a potential game with the finite improvement property, which guarantees our proposed distributed route selection (DRS) scheme will converge to a pure strategy Nash equilibrium within a finite number of iterations.
  Simulation results show that our DRS scheme results in a near-optimal performance that achieves $95\%$ of the maximal total payoff. Moreover, it achieves the highest average payoff and energy efficiency among the benchmark greedy path and circular path schemes.

\end{abstract}

\section{Introduction} \label{sec:introduction}
\subsection{Motivations}

Wireless communication assisted by unmanned aerial vehicles (UAVs) is a promising technology to enhance the performance enhancement of the cellular networks.
 Specifically, UAVs are expected to be deployed in the fifth generation (5G) wireless networks \cite{Tutorial,Survey1,Survey2} by serving as aerial base stations (BSs) or relays to boost the capacity and the coverage of the existing cellular networks \cite{Coverage1,Coverage2,Coverage3,Coverage4}.
 The key reasons for the potential performance enhancement are their high mobility, efficient deployment, and high probabilities of establishing line-of-sight (LOS) connections towards ground users, which improves the quality of service (QoS).
 Thus, UAVs can be deployed to provide Internet coverage to rural areas or cell edges with weak signals, provide extra service capacity for temporary events (such as major sports events and outdoor activities), and restoring communications in emergencies \cite{Tutorial}.

 In recent years, the industry has already started to implement UAV-assisted wireless networks. For example, Google has launched the Loon Project \cite{Loon} with the intention to provide Internet access worldwide.
 To support the deployment of UAVs, the 3rd Generation Partnership Project (3GPP) has studied how the current cellular networks can accommodate UAVs, as well as how to deploy UAV BSs in a convenient and technically feasible manner. For example, its Report in \cite{3GPP} studied how well the existing Long Term Evolution (LTE) radio network can provide services to low-altitude UAVs and the provision of 5G new radio services from high-altitude platforms.
 All these efforts aim to combine UAVs and cellular technologies in a mutually beneficial manner in 5G wireless networks \cite{White}.

 To efficiently provide communication services through UAVs, it is important to consider the \emph{trajectory design} problem regarding \emph{where} and \emph{when} should the operators deploy the UAVs. Prior researches on UAV-enabled wireless communications have mainly focused on the UAVs' trajectory design to improve different QoS requirements\cite{T1,T2,T3,T4,T5,T6,T7,MyGC}.
 The authors in \cite{T1} maximized the minimum throughput over ground users in the downlink communication by optimizing the user scheduling, power control and the UAV's trajectory.
 \cite{T2} investigated the optimal trajectory of UAVs equipped with multiple antennas for maximizing sum-rate in uplink communications.
 The work in \cite{T3} studied the throughput maximization problem in mobile relaying systems by optimizing the source/relay transmit power along with the relay trajectory.
 Moreover, the UAV's transmit power and trajectory were jointly optimized to maximize the minimum average throughput within a given time length in \cite{T4}.
 Furthermore, the authors in \cite{T5} proposed a new cyclical multiple access scheme that the UAV flies cyclically above the ground, and characterized the max-min throughput by optimally allocating the transmission time to ground terminals based on the UAV position.
 In \cite{T6}, a UAV was dispatched to disseminate a common file to a set of ground terminals (GTs) and the authors aimed to design the UAV trajectory to minimize its mission completion time, and \cite{T7} optimized the trajectory design to maximize the amount of energy transferred to all energy receivers during a finite charging period.
 The above studies mainly focus on communication performance improvement and the technical challenges that exist in trajectory design are energy limitation, interference mitigation, and user mobility.

 Since UAVs consume a significant amount of energy to support their mobility, it motivated the design of the \emph{energy-efficient} UAV communication via trajectory design in \cite{E1,E2,E3,E4}.
 The authors of \cite{E1} focused on the energy efficient maximization of a fixed-wing UAV enabled communication for given flight duration.
 The work in \cite{E2} minimizes the total rotary-wing UAV energy consumption while satisfying the individual target communication throughput requirement for multiple ground nodes.
 Moreover, the UAV worked as the mobile data collector has been studied in \cite{E3}, which minimized the maximum energy consumption of all sensor nodes while ensuring that the required amount of data is collected reliably from each sensor node.
 The author in \cite{E4} further studied the trade off between UAV's energy consumption and that of the ground terminals it communicating with.

 Besides energy consumption, another important aspect in the trajectory design is related to the \emph{interference management}.
 As UAVs have high probabilities to establish LOS connections with targets, which will increase the intended signal for the target and also the interference for others, like the ground BSs.
 To address this challenge, current researchers provide a number of techniques, such as the full dimension Multi-input Multi-output (MIMO) multiple antenna BSs, UAVs with directional antennas and beamforming capabilities and power control.
 The authors in \cite{I1} proposed a novel interference-aware path planning scheme for a multi-UAV network to minimize the interference they cause on the ground network.

 More importantly, all of these studies in \cite{T1,T2,T3,T4,T5,T6,T7,MyGC,E1,E2,E3,E4,I1} did not consider the \emph{user mobility} into UAVs' trajectory design.
 In practice, the users may move from one place to another at different times due to their requirements.
 For instance, the stadium may experience communication congestion if a concert holds on it, and if not, it may have fewer user demand.
 If we do not consider the mobility, we will not be able to estimate the demand accurately. As a result, we cannot deploy UAVs to the proper locations at the right times, which reduces the number of users that we can serve.

 \subsection{Contributions}
 In this paper, we maximize the throughput of ground users in the downlink communication by optimizing the UAVs' trajectories, while taking into account the impact of the user mobility, propulsion energy consumption, and UAVs' mutual interference.
 We utilize the Mobility Markov Chains (MMC) \cite{MMC1,MMC2} to model the ground users' movement during a period, so we can estimate the location-and-time dependent user demand.
 The throughput depends on the UAVs' mutual interference.
 Both the user demand and throughput together define the reward function.
 Moreover, the UAV propulsion energy consumption, based on the results in the literature (e.g. \cite{E2}), is defined in the cost function.
 We formulate the problem as a route selection problem in an acyclic directed graph. Each vertex represents a task associated with a reward, and each edge is associated with a cost.
 First, we study the centralized trajectory design, we propose the shortest path scheme that determines the optimal trajectory for the single UAV case.
 We also propose the CRS scheme to systematically compute the optimal trajectories for the more general multiple-UAV case.
 Due to the NP-hardness of the centralized problem, we formulate the UAVs' interactions as a route selection game. We prove that it is a potential game with the FIP, which guarantees our proposed DRS scheme will converge to a pure strategy Nash equilibrium within a finite number of iterations.
 \emph{To the best of our knowledge, this is the first work that brings the user mobility into the UAVs' trajectory design, while taking both the propulsion energy consumption and interference mitigation into account.}

 We summarize the key results and contributions as follows:
 \begin{itemize}
  \item \emph{A general model of UAV trajectory design:} We present a general model of the UAVs' trajectory design with the location-and-time dependent user demand, UAV mutual interference, and propulsion energy consumption.
  \item \emph{Centralized optimal trajectory design:} For the centralized trajectory design, we propose the shortest path scheme for the single UAV case and also the CRS scheme for the multiple-UAV case.
  \item \emph{Distributed route selection algorithm:} We formulate the UAVs' trajectory design as the route selection game, which is a potential game with the FIP. This property guarantees that our proposed DRS scheme will converge to a pure strategy Nash equilibrium within a finite number of iterations.
  \item \emph{Superior performance:} The proposed DRS scheme achieves the best performance in terms of the average payoff and energy efficiency when comparing with various benchmark algorithms, namely the greedy path (GP) scheme and the circular path (CP) scheme \cite{E1,CP}.
 \end{itemize}

The rest of this paper is organized as follows. Section~\ref{sec:system model} describes the system model. Section~\ref{sec:trajectory design} presents the problem formulation and the optimal schemes for the centralized trajectory design. Section~\ref{sec:multiple distributed} shows the DRS scheme for the distributed trajectory design.
Simulation results are provided in Section~\ref{sec:performance evaluation} and Section~\ref{sec:conclusion} concludes this paper.

\section{System Model} \label{sec:system model}
\subsection{UAVs, Regions, and Time Slots}\label{sec:settings}

\begin{figure}[ht]
\centering
       \includegraphics[width=7.55cm, clip = true]{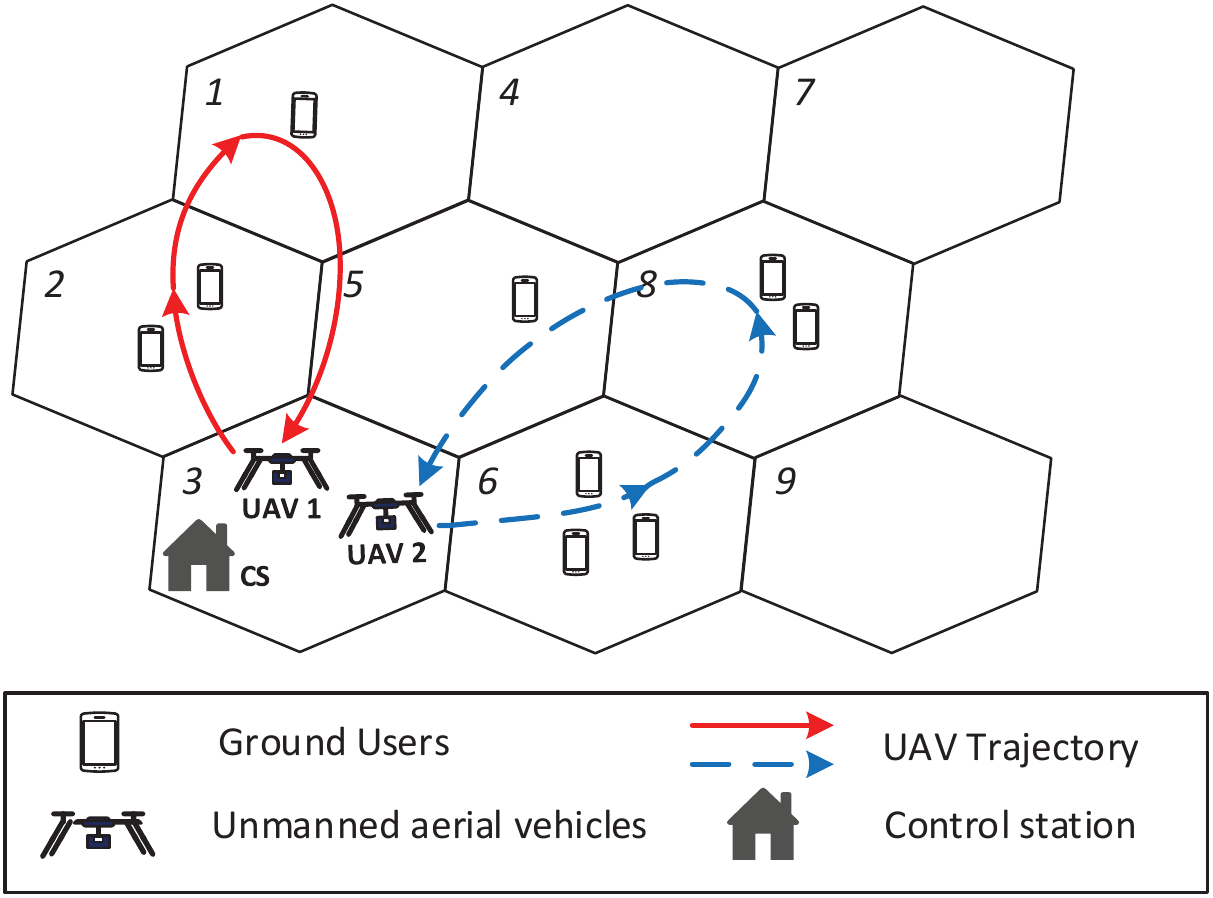}
\caption{An example of the multiple-UAV assisted wireless communication system. Two UAVs start from control station (CS) and provide services for ground users located in $L=9$ regions. The red solid curve represents the trajectory of UAV $1$, which provides communication services first for region $2$ and then region $1$. The blue dashed curve represents the trajectory of UAV $2$, which serves region $6$ and region $8$ away from UAV $1$'s service regions in order to mitigate the mutual interference.} 
   \label{fig:system model graph}
\end{figure}

As shown in Fig.~\ref{fig:system model graph}, we consider a multiple-UAV assisted wireless communication system, where $M \geq 1$ UAVs work as aerial BSs for providing Internet services to ground users located in $L>1$ regions.
The region and UAV sets are denoted as $\mathcal L=\{1,\dots,L\}$ and $\mathcal M=\{1,\dots,M\}$, respectively.
We assume that all the UAVs share the same frequency band for communication over slotted time $t\in \mathcal T=\{1,\dots,T\}$.
Each UAV serves its associated users via multiple access techniques (e.g., TDMA, CDMA, or OFDMA) over a number of orthogonal channels, and it is connected to a nearby macro cell tower with a wireless backhaul link \cite{Backhual}.

We utilize Mobility Markov Chains to characterize the region-and-time dependent user demand, which describes the number of users needed to be served.
Based on the demand (will be discussed in Section~\ref{sec:user demand}), reward and cost of each task, the UAVs have to decide \emph{when} and \emph{where} to provide wireless communication services in the $T$ time slots.
Note that the UAVs start from the control station and need to go back the control station at the end of the serving duration of $T$.
%
\subsection{Tasks Model}\label{sec:task model}
For each region-time point, we define a task. Let $\mathcal K=\{1,\dots,K\}$ be the set of tasks, where $K=LT$.
We map a region-time point $(l,t), l\in\mathcal L, t\in \mathcal T$ to a task index $k \in \mathcal{K}$ by the function
 \begin{equation} \label{equ:mapping function}
     a(l,t)= l + L(t-1).
 \end{equation}
We describe the characteristics related to a task as follows.
 \begin{definition}[Task characteristics] \label{def:task characteristics}
 Each task $k\in \mathcal K$ is associated with:

 \begin{itemize}
   \item The \emph{region} $l_k\in\mathcal L$ and \emph{time slot}\footnote{Each task $k$ is generated at the beginning of the time slot $t_k$. Note that the UAVs must fly to the region $l_k$ before the beginning of the time slot $t_k$ so that it can execute the task. Otherwise, the UAVs cannot work on the task.} $t_k\in \mathcal T$.
   \item The \emph{reward} $\rho_k^m\geq 0$ for UAV $m\in\mathcal M$ completing task $k\in \mathcal K$. The reward is related to other UAVs' decisions (will be defined in Sec.~\ref{sec:reward function}).
   \item The \emph{UAV's potential location}\footnote{Each region is assigned a UAV potential location. When executing task $k$, the UAVs can only stay in this location in region $l_k$.} $\boldsymbol u_{k}=(x_{k}, y_{k}, H)$, where $H$ represents the constant altitude of the UAVs. That is, each UAV can only be in the corresponding location while executing a task.
   \item The \emph{user demand} $\lambda_k\geq 0$, which is the density of the users needed to be served in task $k$ (will be defined in Sec.~\ref{sec:user demand}). We assume that the distribution of the users follows the two-dimensional Poisson Point Process (PPP) $\Phi\in \mathbb R^2$ \cite{PPP}. The users are at the ground level (i.e., zero altitude), so a user's location is denoted by $\boldsymbol v=(x,y,0)$, which is located within the bounds of region $l_k$.
 \end{itemize}
 \end{definition}
To account for the impact of the interference between UAVs on the trajectory design, we will define the reward of each task as the average user throughput subject to the UAVs' mutual interference.
First, we present the user demand and Air-to-Ground (A2G) channel model, on which the reward depends.

\subsection{User Demand}\label{sec:user demand}

In practice, users change their locations over time, and they may belong to different regions.
To properly model the user mobility, we apply the widely adopted Mobility Markov Chains (MMC) \cite{MMC1,MMC2} to estimate the number of the users needed to be served in the $L$ regions, which captures the user demand.

Specifically, we define the probability that a user leaves region $i$ for region $j$ as $p_{ij}$ for $i,j \in \mathcal L$.
Without loss of generality, we further define the region outside $\mathcal L$ as $O$. The user mobility is given by a transmission probability matrix $p=(p_{ij},i,j\in \mathcal R=\mathcal L\cup O)$. Based on MMC, only the current user location is utilized to predict the next one.
We suppose the initial number of users at $t=1$ is $N(l,1)$ for all $l\in \mathcal L$.
Therefore, the expected number of users of region $l$ at time slot $t+1$ is
\begin{equation} \label{equ:MC user number}
    \mathbb{E}[N(l,t+1)]=\mathbb{E}[N(l,t)]+\sum_{j\in \mathcal R \backslash \{l\}}\mathbb{E}[N(j,t)]p_{j,l}-\mathbb{E}[N(l,t)]\sum_{j\in \mathcal R \backslash \{l\}}p_{l ,j},
\end{equation}
where the second term  on the right hand side represents the number of users arriving at region $l$, and the third one is the number of users leaving region $l$. Based on the task model, we define the user demand as $\lambda_k=\lambda_{a(l,t)}=\mathbb{E}[N(l,t)]/S(l)$ (in the unit of number of users per m$^2$), where $S(l)$ is the area of region $l$.

\subsection{Air-to-Ground (A2G) Channel Model}\label{sec:channel model}
\begin{figure}[ht]
\centering
       \includegraphics[width=7.55cm, clip = true]{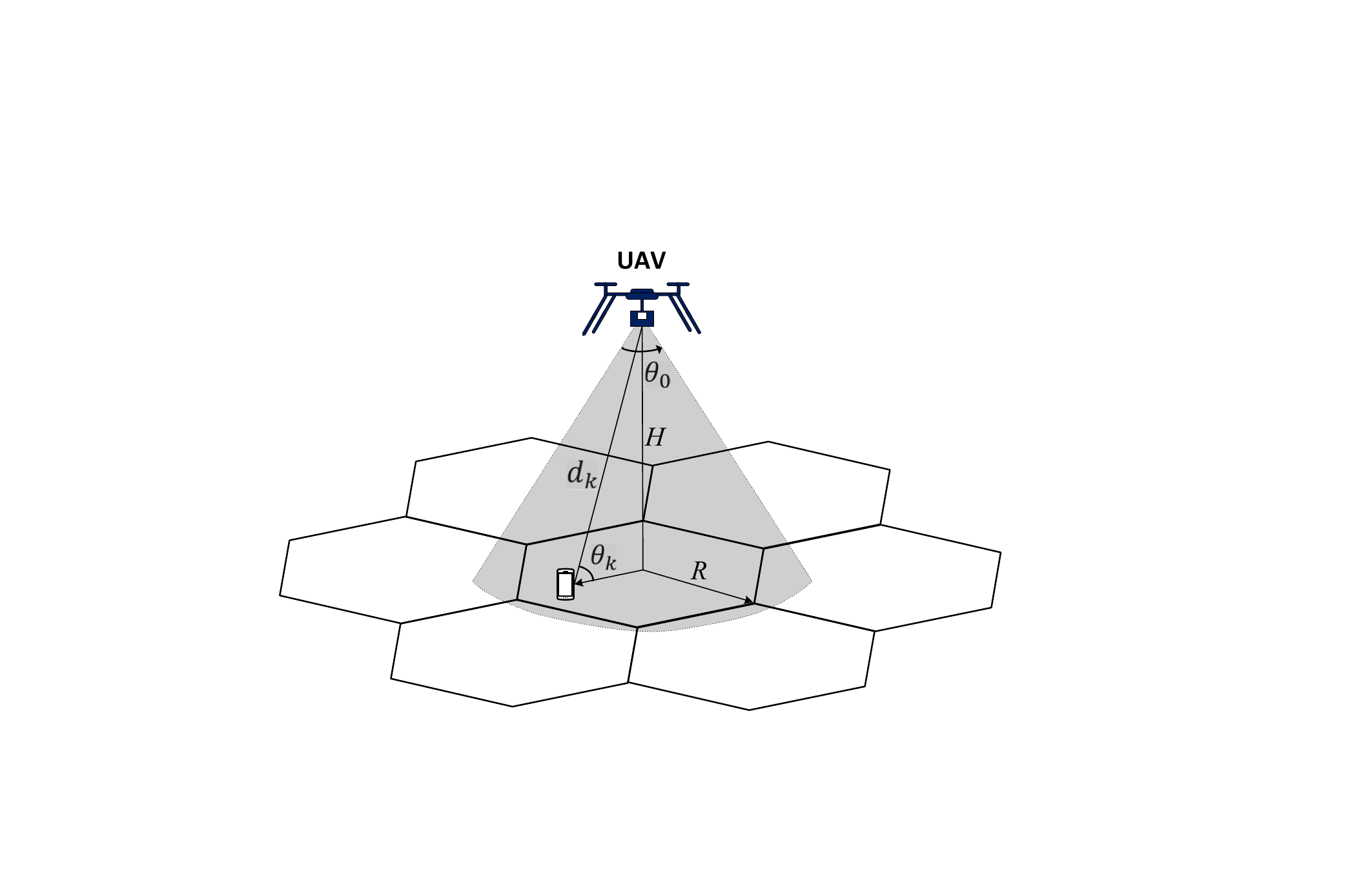}
\caption{Coverage cone of the UAV.} 
   \label{fig:cover_cone}
\end{figure}
%
We assume that UAV $m$ is executing task $k$. Based on the task model in Section~\ref{sec:task model}, the location of UAV $m$ is $\boldsymbol u_{k}=(x_{k}, y_{k}, H)$, and the user's location in region $l_k$ is $\boldsymbol v=(x,y,0)$.
We assume that each UAV has the same type of directional antenna with beamwidth $\theta_0$, and the main beam covers the region directly beneath the UAV.
This coverage cone has a \emph{radius}\footnote{Note that this coverage cone can cover at least one region.} $\tan(\frac {\theta_0}{2})H$.
The interference beyond this coverage is negligible \cite{Backhual,Covercon2}.

As show in Fig.~\ref{fig:cover_cone}, when a user is in the coverage cone ($\theta_k \geq 90^\circ-\frac{\theta_0} {2}$), the path between the UAV and a user can be a LOS path or a non-LOS (NLOS) path.
The LOS probability, which is related to the environment, the user's and UAV's location, and the elevation angle, is given by \cite{LOSP}
\begin{equation} \label{equ:LOS probability}
    P^{LOS}_{k}=\frac 1 {1+\psi \exp(-\zeta[\theta_{k}-\psi])},
\end{equation}
where $\psi$ and $\zeta$ are constant values determined by the type of environment, and $\theta_{k}$ is the elevation angle.
More specifically, $\theta_{k}=\frac {180} \pi \times \sin^{-1}(\frac {H} {d_{k}})$, where $d_{k}=\|\boldsymbol u_{k}-\boldsymbol v\|$ is the Euclidean distance between UAV $m$ and the user located in region $l_k$.
Due to the coverage of the directional antenna, it will not generate interference to neighbouring regions that satisfy $0\leq\theta_k< (90^\circ-\frac{\theta_0}{2})$.
The NLOS probability is $P^{NLOS}_{k}=1-P^{LOS}_{k}$.
The average channel gain \cite{Chanelgain} between the UAV and the user is
\begin{equation} \label{equ:ATG channel gain}
    g_{k}=
 \begin{cases}
\frac {(K_0d_{k})^{-\alpha}} {\eta_1 P^{LOS}_{k}+\eta_2 P^{NLOS}_{k}},& \text{if}\ \theta_k \geq \frac {\theta_0} 2,\\
0,& \text{otherwise},
\end{cases}
\end{equation}
where $K_0=\frac {4\pi f_c} c$, $f_c$ is the carrier frequency, $c$ is the speed of light, and $\alpha$ is the path loss exponent of the link between the UAV and the user. Besides, $\eta_1$ and $\eta_2$ ($\eta_2 > \eta_1>1$) are the excessive path loss coefficients in LOS and NLOS cases.

As all the UAVs share the same frequency band over each time slot, the user may receive the interference from other UAVs.
To avoid severe interference, we assume that each region can only be served by one UAV per time slot.
Let $q_m\in \mathcal L\cup \{0\}$ be the state of UAV $m$, where $q_m = l_k$ means that UAV $m$ is at location $l_k$ executing task $k$, while $q_m = 0$ represents that UAV $m$ is moving to other regions and not executing any task at this moment.
Moreover, we assume that the UAV $m$'s transmission power is $P_m\geq 0$ if it is executing a task (i.e., $q_m \in \mc{L}$) and $P_m=0$ if the UAV is moving (i.e., $q_m = 0$), and will not interfere other communication links.
As a result, the received transmission rate of a user located in region $l_k$ from UAV $m$ at time $t_k$ is
\begin{equation} \label{equ:UAV-User}
    r_{k}^m=\log(1+\frac {P_m g_{k}}{\sum\limits_{\substack{n\neq m, n\in \mathcal M}, \atop q_n\neq 0} P_n g_{a(q_n,t_k)}+N_0}),
\end{equation}
 where $N_0$ is the power of the additive white Gaussian noise.
 In the next subsection, we will define the reward by averaging $r_k^m$ over the position of a random user in region $l_k$.

\subsection{Reward Function}\label{sec:reward function}

Based on \eqref{equ:UAV-User}, we define the \emph{reward} as the average downlink throughput for all the users in the region.
By the PPP assumption, the user locations are independent and identically distributed, so the average throughput can be computed by averaging $r_k^m$ over the region \cite{Avergaethput}.
 Thus, the reward of task $k$ served by UAV $m$ is
\begin{equation} \label{equ:reward function}
\begin{aligned}
    \mathbb \rho_k^m&=\beta \Delta \lambda_k\iint_{D_k} r_k^m{\rm d} x\, {\rm d} y.\\
\end{aligned}
\end{equation}
where $\beta$ denotes the coefficient of the reward, and $\Delta$ represents the fixed bandwidth that each user is allocated under FDMA. Moreover, $D_k=\{(x,y)|\sqrt{(x-x_k)^2+(y-y_k)^2}\leq f(x,y)\}$ is the region corresponding to task $k$. $f(x,y)$ is a function that defines the distance between the boundary and the center of the region.
For example, if we consider a hexagonal topology as in Fig.\ref{fig:system model graph}, it is given by
\begin{equation} \label{equ:regular hexagon}
 f(x,y)=
 \begin{cases}
\frac{\sqrt {3} R}{2\sin(\theta_k^*+\frac {\pi} 3)},& |\theta_k^*|< \frac {\pi} 3,\\
\frac{\sqrt {3} R}{2\sin(\theta_k^*)},& \frac {\pi} 3 \leq |\theta_k^*| < \frac {2\pi} 3,\\
\frac{\sqrt {3} R}{2\sin(\theta_k^*-\frac {\pi} 3)},& \frac {2\pi} 3 \leq |\theta_k^*| \leq \pi,
\end{cases}
\end{equation}
where $\theta_k^*=\arctan (\frac{y-y_k}{x-x_k})$ is the horizontal angle, and $R$ is the side length of each regular hexagon.

\subsection{Cost Function}\label{sec:cost function}

The cost function takes into account both the energy consumption during \emph{flying and hovering}\footnote{Since the communication energy (in the order of a few watts) is usually much smaller than the propulsion energy (in the order of hundreds of watts), we ignore the former in this paper.}.
First, we define the propulsion power consumption of the rotary-wing UAVs\footnote{For the ease of exposition, we assume that all the UAVs adopt the same power consumption model. However, our proposed algorithms in Section~\ref{sec:trajectory design} and~\ref{sec:multiple distributed} can be easily  extended to the more general case with heterogeneous power consumption models.}
 with speed $\phi$ by \cite{E2}
\begin{equation} \label{equ:power consumption}
\begin{aligned}
    P(\phi)=\Lambda_0(1+\frac {3\phi^2}{\omega^2})+\Lambda_1(\sqrt{1+\frac {\phi ^4}{4\chi ^4}}-\frac{\phi^2}{2\chi ^2})^{\frac 1 2}+\frac 1 2 \Lambda_2 \phi ^3,\\
\end{aligned}
\end{equation}
where $\Lambda_0$, $\Lambda_1$, and $\Lambda_2$ are three parameters related to air density and physical properties of the rotor (e.g. rotor solidity and rotor disc area), $\omega$ denotes the tip speed of the rotor blade, and $\chi$ is the mean rotor induced velocity in hovering.

When UAV $m$ flies at a constant speed $\phi_0^m>0$, the power consumption of movement is $P(\phi_0^m)$.
On the other hand, the UAV is hovering and executing a task, it is quasi-stationary when executing a task, so the power consumption of hovering is $P(0)$.

Assume that UAV $m$ aims to perform task $k'\in \mc{K}$ after completing task $k\in \mc{K}$.
The flying distance between these two tasks can be expressed as $d_{k,k'}=\|\boldsymbol u_{k}-\boldsymbol u_{k'}\|$.
Based on the constant speed $\phi_0^m$, the flying duration is $\sigma^m_{k,k'}=\frac {d_{k,k'}}{\phi_0^m}$.
We assume that the length of each time slot is sufficiently small such that the flying duration can be quantized into several time slots, which the length of one time slot is $e$.
The interval between these two tasks is $\xi_{k,k'}=(t_{k'}-t_k-1)e$.
Therefore, task $k$ can be served by UAV $m$ only when $\sigma^m_{k,k'}\leq\xi_{k,k'}$.
For ease of illustration, we define the moving time as
 \begin{equation} \label{equ:moving time}
    \xi_{k,k'}=(t_{k'}-t_k-1)e\geq 0,
 \end{equation}

 By assuming that each task requires one time slot to complete, the hovering time is
 \begin{equation} \label{equ:hovering time}
    \delta _{k,k'}=e.
 \end{equation}
 Overall, we define the cost of UAV $m$ to be the sum of the moving cost and the hovering cost as
\begin{equation} \label{equ:cost function}
    c_{k,k'}^m=\gamma_1 P(\phi_0^m) \xi_{k,k'}+\gamma_2 P(0) \delta _{k,k'} ,
\end{equation}
 where $\gamma _1, \gamma _2\geq 0$ represent the moving and hovering cost per unit watt of the UAV.

\section{Centralized Trajectory Optimization} \label{sec:trajectory design}
 In this section, we focus on the UAVs' trajectory design to maximize their total payoff.
 In Section \ref{sec:graph formulation}, we describe the graph representation of the trajectory design problem.
 In Section~\ref{sec:single}, we study the single UAV trajectory design. We convert the graph so that we can apply the shortest path (SP) algorithm to compute the optimal trajectory.
 Based on the SP scheme, we construct a new graph in Section~\ref{sec:multiple cooperate}, and propose the CRS scheme for multiple UAVs' trajectory design.

%

\subsection{Graph Representation of Trajectory Design} \label{sec:graph formulation}

 Based on the task characteristics in Section~\ref{sec:task model},
 we define the graph associated with regions and time slots as follows for each UAV $m \in \mc{M}$.
 \begin{definition}[Graph representation]\label{def:graph representation}
 Let $\mathcal G_m=(\mathcal V, \mathcal E_m)$ be a graph. We define the set of vertices $\mathcal V$ and the set of edges $\mathcal E_m$ as follows.
 The \emph{vertex set} contains all the region-time points
 \begin{equation} \label{equ:vertex}
   \mathcal V=\{(l,t): l\in \mathcal L, t\in \mathcal T\}.
 \end{equation}
Each region-time point is associated with a \emph{reward} $\mathbb \rho_{a(l,t)}^m$ defined in \eqref{equ:reward function}.
The \emph{edge set} contains the feasible transitions of UAV $m$ between any two different region-time points as
 \begin{equation} \label{equ:edge}
 \begin{split}
 \mathcal E_m=\{((l,t),(l',t')): l,l'\in \mathcal L, t,t'\in \mathcal T, t' > t, \sigma^m_{a(l,t),a(l',t')}\leq\xi_{a(l,t),a(l',t')}\}.
 \end{split}
 \end{equation}
 From Section \ref{sec:cost function}, an edge $((l,t),(l',t')) \in \mc{E}_m$ exists only when UAV $m$ can arrive in region $l'$ at time slot $t'$.
 Each edge is associated with the \emph{cost} $c_{a(l,t),a(l',t')}^m$ defined in \eqref{equ:cost function}.
 \end{definition}

 Next, we formulate UAV $m$'s trajectory design as the region-time route selection problem in graph $\mathcal G_m$. We define the feasible routes for UAV $m$ as follows.
 \begin{definition}[Feasible route] \label{def:feasible route}
 Based on graph $\mathcal G_m$, feasible route of UAV $m$ is
 \begin{equation} \label{equ:route}
     s_m =((l^1_m,t^1_m), (l^2_m,t^2_m), \ldots, (l^n_m,t^n_m))\in \mathcal S_m,
 \end{equation}
 where the vertex $(l^i_m,t^i_m)\in \mathcal V, \forall i=1,\dots,n $, and the edge $((l^i_m,t^i_m),(l^{i+1}_m,t^{i+1}_m))\in \mathcal E, \forall i= 1,\dots,n-1$.
 For the first region-time point $(l^1_m, t^1_m)$, $l^1_m$ represents the source and $t^1_m = 1$. For the last region-time point $(l^n_m, t^n_m)$, $l^n_m$ represents the destination and $t^n_m = T$.
 \end{definition}

\begin{figure}[ht]
\centering
       \includegraphics[width=7.55cm, trim = 0cm 1.3cm 0cm 0cm, clip = true]{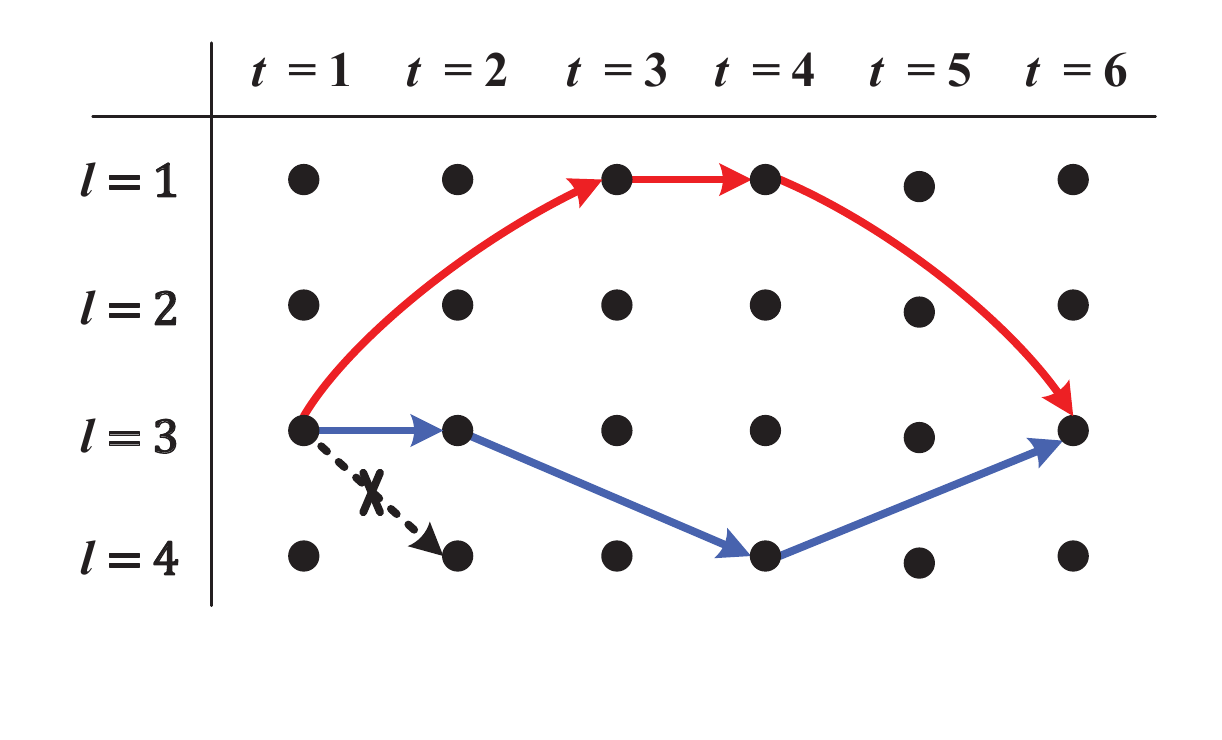}
\caption{Some examples of the UAV's possible trajectories. The red solid line that represents the feasible route of UAV $1$ is $s_1=((3,1),(1,3),(1,4),(3,6))$, and the blue solid line is the feasible route of UAV $2$, which is $s_2=((3,1),(3,2),(4,4),(3,6))$.} 
   \label{fig:graph G}
\end{figure}
\begin{example}
{
 We show an example in Fig.~\ref{fig:graph G}. We can see that the edge $((3,1),(4,2))$ is not feasible, because the UAV cannot arrive in region $4$ before the beginning of the second time slot.
 Besides, we have two feasible routes for UAV $1$ and UAV $2$, which are $s_1=((3,1),(1,3),(1,4),(3,6))$ and $s_2=((3,1),(3,2),(4,4),(3,6))$, respectively. In this example, the source and destination are both in the \emph{same}\footnote{Note that the scheme we propose in Sections~\ref{sec:single}, ~\ref{sec:multiple cooperate} and~\ref{sec:multiple distributed} are suitable for an arbitrary source and destination.} region (e.g. region $3$).
 }
 \end{example}

 We define $\mathcal V(s_m)$ and $\mathcal E(s_m)$ as the set of vertices and edges traversed by the route $s_m$.
 For example, the set of vertices of route $s_2$ is $\mathcal V(s_1)=\{(3,1),(1,3),(1,4),(3,6)\}$, and the set of edges is $\mathcal E(s_m)=\{((3,1),(1,3)),((1,3),(1,4)),((1,4),(3,6))\}$.
 The strategy profile $\boldsymbol s = (s_1,\dots, s_M) \in \mathcal S_1\times\dots\times \mathcal S_M$ is the strategies of all the UAVs.
 Therefore, the payoff (i.e., total rewards minus total costs) that UAV $m$ gets for choosing route $s_m$ in a strategy profile $\boldsymbol s$ is to
\begin{equation} \label{equ:each UAV payoff}
 \begin{split}
 U_m(\boldsymbol s)=\sum_{(l,t)\in\mathcal V(s_m)} {\rho} _{a(l,t)}^m - \sum _{((l,t),(l',t'))\in \mathcal E(s_m)}c_{a(l,t),a(l',t')}^m.
 \end{split}
\end{equation}
The goal of the centralized trajectory optimization is to maximize the UAVs' total payoff as
\begin{equation} \label{equ:total payoff}
 \begin{split}
 \max_{\boldsymbol s} \quad \sum _{m\in \mathcal M} U_m(\boldsymbol s),
 \end{split}
\end{equation}
which is an NP-hard problem. We first propose the shortest path (SP) scheme for the single UAV trajectory design in Section \ref{sec:single}, which serves as the basis for the more general multiple UAVs' optimal trajectory design in Section \ref{sec:multiple cooperate}.

\subsection{Single UAV Trajectory Design} \label{sec:single}

For the single UAV trajectory design (i.e., $M=1$), there is no interference caused by other UAVs.
According to \eqref{equ:reward function} and \eqref{equ:cost function}, the reward and the cost can be pre-determined.
Based on graph $\mathcal G_m$ in Fig.~\ref{fig:graph G}, we suppose the red and blue solid line represent two feasible routes for the single UAV.
Since each vertex is associated with a reward and each edge is associated with a cost, graph $\mathcal G_m$ is not a \emph{standard acyclic directed graph}\footnote{A standard acyclic directed graph is formed by a collection of vertices and edges, where the vertices are structureless objects that are connected in pairs by edges. Each edge has an orientation, from one vertex to another vertex, and is associated with a weight. However, each vertex is not associated with any weight \cite{Alg}.}.
To handle this problem, we propose the shortest path (SP) scheme as shown in Scheme \ref{Alg:1}.

We first convert it into a new graph $\mathcal G_{m}^*$, and apply the \emph{Bellman-Ford algorithm}\footnote{We use Bellman-Ford algorithm, because it can compute the shortest path on a graph with both positive and negative edge weight. We can use the Dijkstra Algorithm for our problem with the lower time complexity, but it needs one more step conversion.} to find the \emph{optimal} route in problem \eqref{equ:total payoff}, which takes the following steps.
\begin{figure}[t]
\centering
       \includegraphics[width=6.0cm, clip = true]{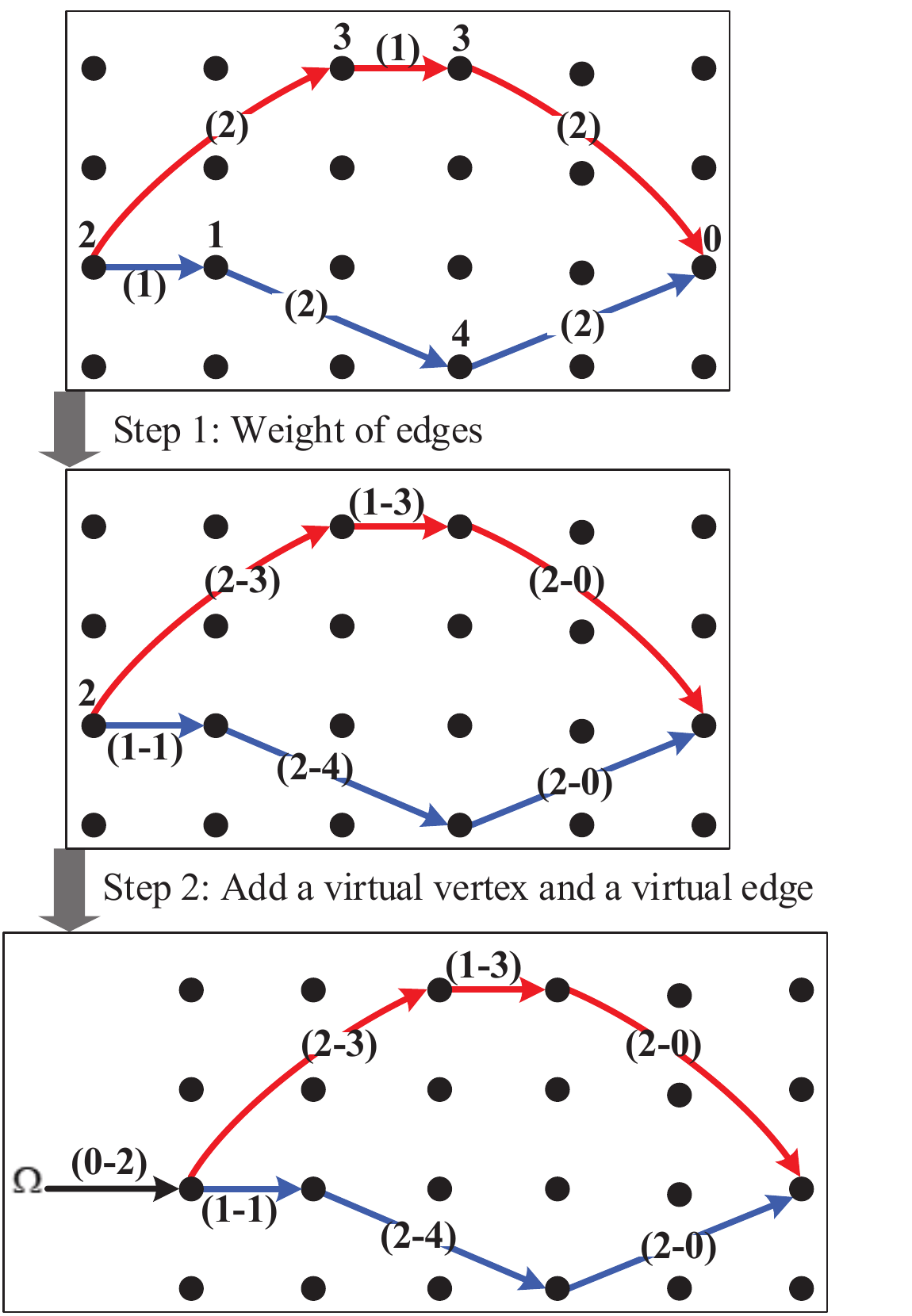}
\caption{An illustration for the graph conversion from graph $\mathcal G_m$ to graph $\mathcal G_{m}^*$. Each vertex represents one task and is associated with a reward (the number without a bracket). Each edge is associated with a cost (the number with a bracket). The red and blue solid line represent two feasible routes for the single UAV.}
   \label{fig:graph conversion}
\end{figure}
\begin{itemize}
    \item [1)] \textit{The weight of edges:} The newly defined edges incorporate both the rewards and the costs, where the weight of an edge in graph $\mathcal G_{m}^*$ as
        \begin{equation}
        w_{a(l,t),a(l',t')}^m = c_{a(l,t),a(l',t')}^m-\rho_{a(l',t')}^m.
        \end{equation}
        In this way, we will not associate any rewards with the vertices.
    \item [2)] \textit{Add a virtual vertex and a virtual edge to graph $\mathcal{G}_{m}^*$:} The UAV sets off from the control station.
        We have the reward of the starting point ${\rho}_{a(l^1_m,t^1_m)}^m$ defined in \eqref{equ:route}.
        We use $\Omega$ to represent the virtual vertex, and add a virtual edge to connect it with the starting point (see the third subfigure in Fig.~\ref{fig:graph conversion}).
        Therefore, the weight of this virtual edge is $-\rho_{a(l^1_m,t^1_m)}^m$. We use $X$ to represent this virtual edge.
        Fig.~\ref{fig:graph conversion} illustrates how these two steps convert graph $\mathcal G_m$ into graph $\mathcal{G}_{m}^*$.
    \item [3)] \emph{Run Bellman-Ford Algorithm:} Based on graph $\mathcal G_{m}^*$, the Bellman-Ford algorithm \cite{Alg} is applied  from source $\Omega$ to destination $(l_m^{n}, T)$ (see Definition~\ref{def:feasible route}) to find the optimal route $\boldsymbol s^*$.
    \item [4)] \textit{Conversion from cost minimization to payoff maximization:} The minimal cost computed by the Bellman-Ford algorithm is opposite to the maximum payoff.
        That is,
        \begin{equation}
        U(s_m^*)=-\sum \limits_{((l,t),(l',t'))\in \mathcal E(s_m^*)\cup X}w_{a(l,t),a(l',t')}^m.
        \end{equation}
\end{itemize}

\begin{algorithm}[t]
  \caption{Shortest Path (SP) Algorithm}    \label{Alg:1}
  \KwIn{Vertex set $\mc{V}$ in \eqref{equ:vertex}; Vertex reward $\rho_{a(l,t)}^m, \forall (l,t) \in \mc{V}$ in \eqref{equ:reward function}; Edge set $\mc{E}_m$ in \eqref{equ:edge}; Edge cost $c_{a(l,t), a(l',t')}^m, \forall ((l,t),(l't')) \in \mc{E}_m$ in \eqref{equ:cost function};}
  Compute the weights of graph $\mathcal G^*_m$: $w_{a(l,t),a(l',t')}^m = c_{a(l,t),a(l',t')}^m-\rho_{a(l',t')}^m, \ \forall ((l,t),(l't')) \in \mc{E}_m$\;
  Add a virtual vertex $\Omega$ and a virtual edge $X$\;
  Run Bellman-Ford Algorithm in graph $\mc{G}_m^*$ from source $\Omega$ to destination $(l_m^{n}, T)$ for the optimal route $s_m^*$\;
  Calculate payoff $U(s_m^*)=-\sum \limits_{((l,t),(l',t'))\in \mathcal E(s_m^*)\cup X}w_{a(l,t),a(l',t')}^m$\;
    \KwOut{Route $s_m^*$.}
\end{algorithm}
 \vspace{0.2cm}
 \begin{proposition} \label{PRO:time complexity}
   The UAV can determine its optimal route $\boldsymbol s$ within $\mathcal O(L^3T^2)$ time.
 \end{proposition}
 \vspace{0.2cm}
 \begin{proof} \label{PROF:time complexity}
 The Bellman-Ford Algorithm has a time complexity $\mc O(VE)$ \cite{Alg}, where $V$ and $E$ represent the number of vertices and edges, respectively.
 For graph $\mathcal G^*_m$, we have $V=LT$ and $E=\sum\limits_{i=1}^{T}L^2(T-i)$.
 Therefore, the SP scheme has a time complexity $\mathcal O(L^3T^2)$.
 \end{proof}
 \vspace{0.2cm}
%

\subsection{Multiple UAVs' Trajectory Design} \label{sec:multiple cooperate}

For the multiple UAVs' trajectory design (i.e., $M>1$), the reward of each task cannot be pre-determined, because it is related to the states of all the UAVs.
Therefore, we cannot obtain graph $\mathcal G_m$ directly in this scenario.

As shown in Fig.~\ref{fig:graph GMstar}, we first construct a new graph $\mathcal F=(\mathcal A,\mathcal J)$ by considering the \emph{state vector} $\boldsymbol q=(q_1,\dots,q_M), q_m\in \mathcal L\cup\{0\}$, which is defined in Section~\ref{sec:channel model}. Let $\mathcal Q=\{\boldsymbol q_1,\dots,\boldsymbol q_B\}$ be the set of all possible state vectors, where $B=\sum_{m=0}^{M}\mathrm C_M^m \frac{N!}{(N-(M-m))!}$ represents the total number of state vectors \cite{Combi}.
Similar to the discussion in Section~\ref{sec:single}, we will define the vertex set, vertex reward, edge set, and edge cost of the new graph $\mc F$ and compute the optimal trajectories.

\subsubsection{Vertex set and rewards of the new graph}
Each vertex is represented by a state-time point $(\boldsymbol q,t)$, and we define the vertex set $\mathcal A$ as
 \begin{equation} \label{equ:new vertex}
   \mathcal A=\{(\boldsymbol q,t): \boldsymbol q \in \mathcal Q, t\in \mathcal T\}.
 \end{equation}
The \emph{vertex reward} of vertex $(\boldsymbol q,t)$ is the summation of each UAV's reward in \eqref{equ:reward function}:
\begin{equation} \label{equ:Mutiple reward}
 \begin{split}
 \Upsilon(\boldsymbol q,t)=\sum\limits_{m\in \mathcal M}\rho^m_{a(q_m,t)}.
 \end{split}
\end{equation}

\subsubsection{Edge set and costs of the new graph}
Each edge indicates changes in the UAVs' states. Before calculating the associated cost, we first analyze the condition for a feasible edge.
For each UAV, we need to record the starting region and the number of time slots it has been moving, so that we can decide whether the edge exists or not. More specifically, we define a \emph{mobility vector} for each vertex $(\boldsymbol q,t)$ as $\boldsymbol I=(I_1,\dots,I_M)$, which decides whether the UAVs are moving at time slot $t$. The element of mobility vector is
\begin{equation} \label{equ:mobility sequence}
  I_m= \textbf {1}_{\{q_m=0\}}, q_m\in\boldsymbol q,
\end{equation}
where $\textbf {1}_{\{.\}}$ is the indicator function.
We further define the state vector $\overline {\boldsymbol q}=(q_1,\dots,q_M)$ for each vertex $(\boldsymbol q,t)$, which records the starting region for each UAV.
Moreover, we define the updating mobility vector $\overline {\boldsymbol I}=(I_1\dots,I_M)$ for $\boldsymbol I$ to record how many time slots that each UAV is moving.
We consider that $((\boldsymbol q,t),(\boldsymbol q',t+1))$ is an edge connected vertex $(\boldsymbol q,t)$ and $(\boldsymbol q',t+1))$.
We update the state vector and the updating mobility vector for vertex $(\boldsymbol q',t+1)$ as
%
\begin{equation} \label{equ:update status}
 \overline q_m'=
 \begin{cases}
q_m,& \text{if } q_m'=0,\\
q_m',& \text{otherwise}.
\end{cases}
\end{equation}
%
%
\begin{equation} \label{equ:update mobility}
 \overline I_m'=
 \begin{cases}
I_m',& \text{if } I_m'=0,\\
I_m'+I_m,& \text{otherwise},
\end{cases}
\end{equation}
which records how many time slots each UAV is moving until the current time.
According to Section~\ref{sec:cost function}, the condition for the feasible edges is
\begin{equation} \label{equ:feasible condition}
\sigma^m_{a(\overline {q}_m,t),a(\overline {q}_m',t+1)}\leq\xi_{a(\overline {q}_m,t),a(\overline {q}_m',t+1)}+{I}_me, \quad \forall \overline {q}_m\in\boldsymbol{\overline q}, \overline {q}_m'\in\boldsymbol{\overline q}', {I}_m\in \boldsymbol I.
\end{equation}
It indicates that each UAV has to arrive in the corresponding regions of state $\boldsymbol {\overline q}'$ at the beginning of time slot $t+1$.

We define the set of edges as
 \begin{equation} \label{equ:new edge}
 \begin{split}
 \mathcal J=\{((\boldsymbol q,t),(\boldsymbol q',t')): \boldsymbol q,\boldsymbol q'\in \mathcal Q, t,t'\in \mathcal T, t' > t, \sigma^m_{a(\overline {q}_m,t),a(\overline {q}_m',t+1)}\leq\\\xi_{a(\overline {q}_m,t),a(\overline {q}_m',t+1)}+{I}_me, m\in \mathcal M\}.
 \end{split}
 \end{equation}

Based on the aforementioned discussion, some UAVs' states may be equal to $0$, which means they are moving for one time slot.
On the other hand, some UAVs may be hovering for one time slot to serve the corresponding regions.
Therefore, the \emph{edge cost} for $((\boldsymbol q,t),(\boldsymbol q',t+1))$ is
\begin{equation} \label{equ:Mutiple cost}
 \begin{split}
 \Phi(\boldsymbol q,\boldsymbol q')=\sum_{q_m\in \boldsymbol q'}\gamma_1^m(\textbf {1}_{\{q_m=0\}}e)+\sum_{q_m\in \boldsymbol q'}\gamma_2^m(1-\textbf {1}_{\{q_m=0\}})e.
 \end{split}
\end{equation}
\begin{figure}[t]
\centering
       \includegraphics[width=9.5cm, trim = 0cm 0.1cm 0cm 0cm, clip = true]{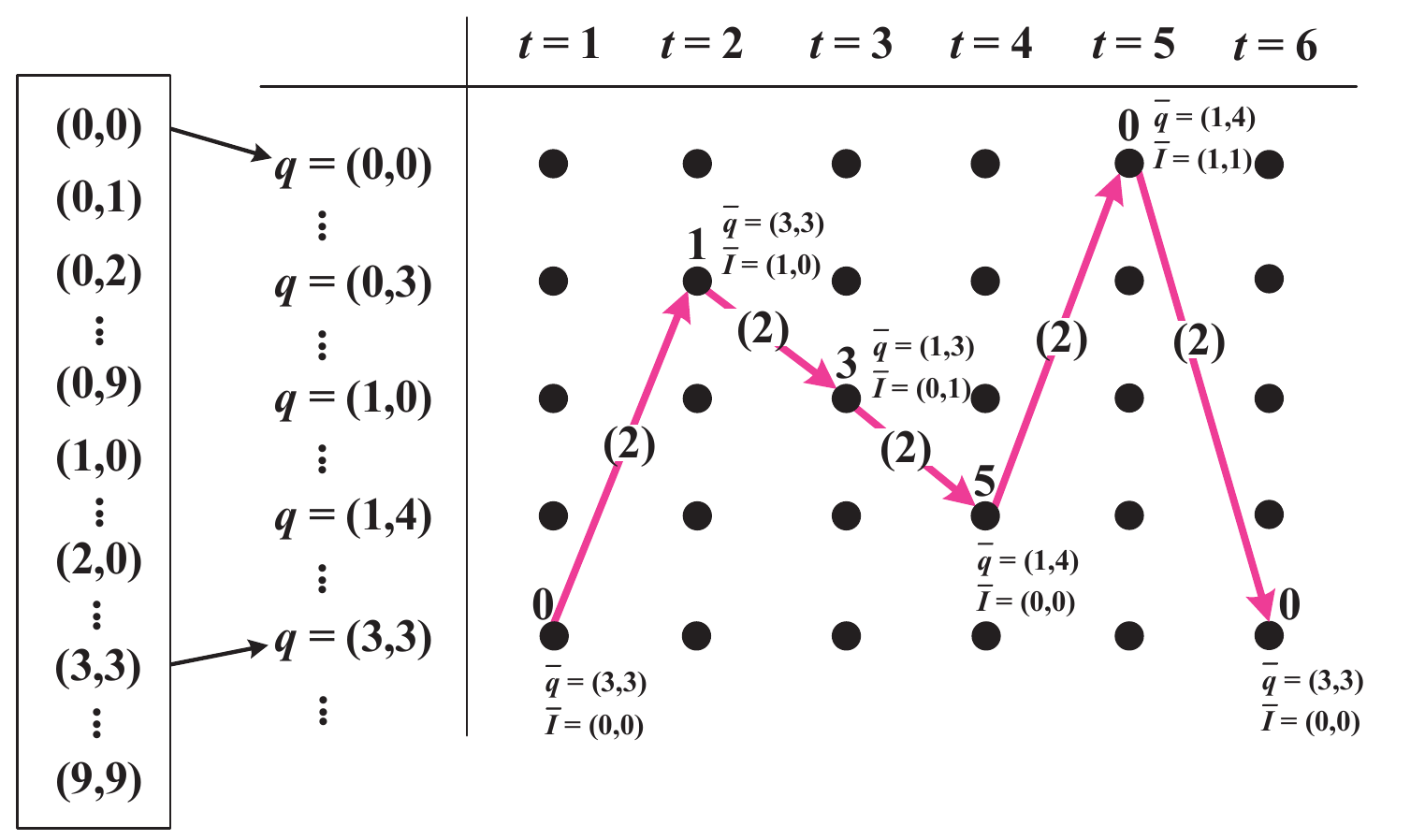}
\caption{An example graph $\mathcal F$ of multiple UAV trajectory design. The route of UAV $1$ and UAV $2$ are $((3,1),(0,2),(1,3),(1,4),(0,5),(3,6))$ and $((3,1),(3,2),(0,3),(4,4),(0,5),(3,6))$, where $0$ means the UAV is moving for one time slot.}
   \label{fig:graph GMstar}
\end{figure}

\begin{example}
{
In the pink solid route in Fig.~\ref{fig:graph GMstar}, UAV $1$ and UAV $2$ are serving the same region $3$ at $t=1$. The state vector is $\overline{\boldsymbol q}=(3,3)$, and the updating mobility vector is $\overline{\boldsymbol I}=(0,0)$ because both UAVs are providing services. At $t=2$, UAV $1$ is moving, and UAV $2$ is still serving region $3$. Therefore, the state vector, which records the starting region for all the UAVs, is $\overline{\boldsymbol q}=(3,3)$. The updating mobility vector is $\overline{\boldsymbol I}=(1,0)$, which records UAV $1$ is moving for one time slot. Based on the movement speed and distance, we have that UAV $1$ can arrive at region $1$ after flying for one time slot.
According to \eqref{equ:feasible condition}, the second edge is feasible.
Accordingly, the mobility vector and state vector for each vertex in the route can be computed.
Based on these two vectors, we can be computed the pre-determined cost for each feasible edge by \eqref{equ:Mutiple cost}.
In graph $\mathcal F$, each vertex is associated with a reward in \eqref{equ:Mutiple reward}, and each edge is associated with a cost in \eqref{equ:Mutiple cost}.
Furthermore, the route of UAV $1$ and UAV $2$ are $((3,1),(0,2),(1,3),(1,4),(0,5),(3,6))$ and $((3,1),(3,2),(0,3),(4,4),(0,5),(3,6))$.
}
\end{example}

 \subsubsection{Feasible route in graph $\mc F$}
 Based on graph $\mathcal F$, the feasible route of all the UAVs is
 \begin{equation} \label{equ:new route}
     \overline {s} =((\boldsymbol q^1,1), (\boldsymbol q^2,2), \ldots, (\boldsymbol q^T,T))\in \overline {\mathcal S},
 \end{equation}
 where $\overline {\mathcal S}$ represents the feasible route set for the UAVs in graph $\mathcal F$. Also, we have the vertex $(\boldsymbol q^i,i)\in \mathcal A, \forall i=1,\dots,n $, and the edge $((\boldsymbol q^i,i),(\boldsymbol q^{i+1},{i+1}))\in \mathcal J, \forall i= 1,\dots,T-1$.
 For the first state-time point $(\boldsymbol q^1, 1)$, $\boldsymbol q^1$ represents the sources of all the UAVs.
 For the last state-time point $(\boldsymbol q^T, T)$, $\boldsymbol q^T$ represents the destinations of all the UAVs.

 \subsubsection{Multiple UAVs' optimal trajectories}
 We define $\mathcal A(\overline {s})$ and $\mathcal J(\overline {s})$ as the set of vertices and edges traversed by the route $\overline {s}$.
 Therefore, the payoff $U(\overline {s})$ that all the UAVs get for choosing route $\overline {s}$ is equal to
\begin{equation} \label{equ:new total payoff}
 \begin{split}
 U( \overline {s})=\sum_{(\boldsymbol q, t)\in\mathcal A(\overline {s})} \Upsilon(\boldsymbol q,t) - \sum _{((\boldsymbol q,t),(\boldsymbol q',t+1))\in \mathcal J(\overline {s}))}\Phi(\boldsymbol q,\boldsymbol q'),
 \end{split}
\end{equation}
so that we can convert the original problem \eqref{equ:total payoff} into
\begin{equation} \label{equ:new total payoff conversion}
 \begin{split}
 \max_{\overline{\boldsymbol s}} \quad \sum _{m\in \mathcal M} U(\overline{\boldsymbol s}).
 \end{split}
\end{equation}

In the graph $\mathcal F$,  with vertex set $\mc A$ in \eqref{equ:new vertex}, vertex reward $\Upsilon(\bs{q},t), \forall (\bs{q},t) \in\mc A$ in \eqref{equ:Mutiple reward}, edge set $\mc J$ in \eqref{equ:new edge}, and edge cost $\Phi(\bs{q},\bs{q}'), \forall \bs{q},\bs{q}' \in \mc{Q}$ in \eqref{equ:Mutiple cost} as input, we can adopts the ideas in the SP scheme in Algorithm \ref{Alg:1} to compute the optimal trajectories for multiple UAVs.

 \vspace{0.2cm}
 \begin{proposition} \label{PRO:time complexity for multiple}
   Based on the SP scheme, multiple UAVs can determine their optimal route $\overline s$ within $\mathcal O(N^{3M}T^2)$ time.
 \end{proposition}
 \vspace{0.2cm}
 \begin{proof} \label{PROF:time complexity for multiple}
   The idea is similar to the proof of Proposition \ref{PRO:time complexity} with $V = BT$ and $E = B^2(T-1)$ in graph $\mc F$.
 \end{proof}
 \vspace{0.2cm}
%

Multiple UAVs' centralized trajectory design is an NP-hard problem, which is computationally complex to solve for large networks.
Besides, it requires a central entity with the full knowledge of the current state of the network and the ability to communicate with all
UAVs at all time. However, this might not be feasible in case the UAVs belong to different operators or in scenarios in which the environment changes dynamically. Therefore, we next adopt a distributed trajectory design in which each UAV decides autonomously on its trajectory.

\section{Distributed Route Selection Game} \label{sec:multiple distributed}

In this section, we first formulate the UAVs' trajectory design problem as the Route Selection Game (RSG), and then propose the distributed route selection algorithm (DRS).
Our objective is to develop a distributed approach that allows each UAV to choose its trajectory in an autonomous manner.

\subsection{Non-cooperative Route Selection Game}
To avoid multiple UAVs from serving in the same region, which is inefficient due to their severe mutual interference, we define the reward of task $k$ as
\begin{equation} \label{equ:reward distributed}
 \varrho_{k}(z_k)=
 \begin{cases}
 \beta \Delta \lambda_k\iint_{D_k} \log(1+\frac {P_m g_{k}}{N_0}){\rm d} x\, {\rm d} y,& z_k=1,\\
 0,& z_k=0 \ \text{or} > 1,
\end{cases}
\end{equation}
where $z_k$ is the number of UAVs serving task $k$. The first line is the reward of task $k$ served by UAV $m$ in \eqref{equ:reward function} when there is no interference caused by other UAVs.
Since each task can only be served by one UAV, we define that the reward of task $k$ will be $0$ if multiple UAVs serve the same task.
According to \eqref{equ:each UAV payoff}, the payoff of UAV $m$ with route $s_m$ is
\begin{equation} \label{equ:payoff distributed}
 \begin{split}
 \widetilde{U}_m(\boldsymbol s)=\sum_{(l,t)\in\mathcal V(s_m)} {\varrho} _{a(l,t)}(z_{a(l,t)}(\bs s)) - \sum _{((l,t),(l',t'))\in \mathcal E(s_m)}c_{a(l,t),a(l',t')}^m,
 \end{split}
\end{equation}
where $z_{a(l,t)}(\bs s)$ is the number of UAVs providing services for task $k=a(l,t)$ based on the current route $\bs s$.

Based on the graph representation in Section~\ref{sec:graph formulation}, we formulate the multiple UAV trajectory design as a RSG, where UAVs act as players to choose the available region-time routes.
 \begin{definition}[Route selection game]
 A route selection game is a tuple $\Gamma=(\mathcal M, \mathcal S,(\widetilde {U}_m)_{m\in \mathcal M})$ defined by:
 \begin{itemize}
   \item Players: The set of UAVs $\mc{M}$.
   \item Strategies: The set of routes of all the UAVs is $\mc{S}$ = $\mc{S}_1 \times \ldots\mc{S}_M$ defined in \eqref{equ:route}.
   \item Payoffs: $(\widetilde {U}_m)_{m\in \mathcal M}$ contains the payoff functions of the UAVs defined in \eqref{equ:payoff distributed}.
 \end{itemize}
 \end{definition}

Let $\boldsymbol s_{-m}=(s_1,\dots,s_{m-1},s_{m+1},\dots,s_M)$ be the strategies of all the UAVs except UAV $m$. For the RSG, we are interested in whether the UAVs can reach an equilibrium strategy profile $\boldsymbol s=(s_m, \boldsymbol s_{-m})$, in which no UAV can further increase its own payoff by changing its strategy, i.e., a pure Nash equilibrium of the game $\Gamma$.
 \begin{definition}[Pure Nash equilibrium]
 A Nash equilibrium (NE) \cite{Potentialgame} of the game $\Gamma$ is a strategy profile $\boldsymbol s^*$ such that $$\widetilde U_m(s^*_m, \boldsymbol s^*_{-m})\geq \widetilde U_m(s_m,\boldsymbol s^*_{-m}), \forall s_m\in \mathcal S_m, \forall m\in \mathcal M.$$
 \end{definition}

\begin{definition}[Best response update]
Given a strategy profile $(s_m, \boldsymbol s_{-m})$, we say that strategy $s_m'$ is a best response update for UAV $m$ if it maximizes UAV $m$'s payoff. That is, $s_m' = \arg\max\limits_{s_m \in \mc{S}_m} \tilde{U}(s_m, \bs{s}_{-m})$.
\end{definition}
 \begin{definition}[Finite improvement property]
 A game possesses the finite improvement property (FIP) \cite{Potentialgame} when each UAV's best response update always converge to a pure NE within a finite number of steps, irrespective of the initial strategy profile or the updating order of the UAVs.
 \end{definition}

The FIP implies that best response updating always lead to a pure NE, which implies the \emph{existence} of a pure NE \cite{MichaelMBHOC,MichaelArticle}.
 \vspace{0.2cm}
 \begin{theorem} \label{Thm:FIP}
   Every route selection game possesses the FIP.
 \end{theorem}
 \vspace{0.2cm}

The proof of Theorem \ref{Thm:FIP} is given in Appendix A.
%


%
\begin{algorithm}[t]
  \caption{Distributed Route Selection Algorithm (DRS)}
  \label{Alg:DRS}
  \KwIn{Task set $\mathcal K$; UAV set $\mathcal M$; user demand $\lambda_k, k\in \mathcal K$; speed of the UAV $\phi_0^m, m\in\mathcal M$; region set $\mathcal L$; time set $\mathcal T$; UAV potential location for each task $\boldsymbol u_k, k\in \mathcal K$; UAV transmission power $P_m, m\in \mc M$; region side length $R$;}
  According to Definition~\ref{def:graph representation}, prepare the vertex set $\mc{V}$ in \eqref{equ:vertex} and edge set $\mc{E}_m, \forall m\in\mc M$ in \eqref{equ:edge}\;
  Calculate the edge cost $c_{a(l,t), a(l',t')}^m, \forall ((l,t),(l't')) \in \mc{E}_m, m\in \mc M$ in \eqref{equ:cost function}\;
  Based on Definition~\ref{def:feasible route}, find the feasible route set for all the UAVs $\mathcal S_m, \forall m\in \mathcal M$\;
  Initialize the routes for all the UAVs $\boldsymbol s:=(s_1,\dots,s_M), s_m\in\mc S_m, \forall m\in \mc M$\;
  \Repeat
  {\rm {\text{The strategy of each UAV does not change. That is, $s_m'=s_m^*, \forall m\in\mc M$}}}
  {
  \For {\textbf{\rm{each}} $m\in \mathcal M$}
  {
    Calculate the number of serving UAVs for all tasks $ k \in \mathcal K:z_k=\sum\limits_{m=1}^{M}\textbf {1}_{\{(l_k,t_k)\in s_m\}}$\;
    Compute the vertex reward ${\varrho} _{k}(z_{k}), \forall k\in \mc K$ in \eqref{equ:reward distributed}\;
    Derive UAV $m$'s payoff:
    \begin{equation}\label{equ:algorithm2utility}
    \widetilde{U}_m(s_m, \boldsymbol s_{-m})=\sum_{(l,t)\in\mathcal V(s_m)} {\varrho} _{a(l,t)}(z_{a(l,t)}) - \sum _{((l,t),(l',t'))\in \mathcal E(s_m)}c_{a(l,t),a(l',t')}^m;
    \end{equation}
    Update $z_{a(l,t)}:=z_{a(l,t)}+1, \forall (l,t)\in \mathcal V(\boldsymbol s_{-m})\backslash\mathcal V(s_{m})$\;
    Based on \eqref{equ:reward distributed}, update the vertex reward ${\varrho} _{k}(z_{k}), \forall k\in \mc K$\;
    Run Algorithm \ref{Alg:1} (shortest path (SP) scheme) to obtain the optimal route $s_m'=s_m^*\in \mathcal S_m$\;
    Based on \eqref{equ:algorithm2utility}, update UAV $m$'s payoff $\widetilde{U}_m(s_m, \boldsymbol s_{-m})$\;
    \If{$\widetilde{U}_m(s_m', \boldsymbol s_{-m})-\widetilde{U}_m(s_m, \boldsymbol s_{-m}) > 0$}
      {
        Update the strategy profile $\boldsymbol s:=(s_m',\boldsymbol s_{-m})$\;
      }
  }
  }
    \KwOut{Strategy profile: $\boldsymbol s$.}
\end{algorithm}

\subsection{Distributed Route Selection Algorithm}
Based on the FIP, we propose a distributed route selection algorithm (DRS) as shown in Algorithm \ref{Alg:DRS}, and which consists of the following steps.

\emph{Initialization:} According to Definition~\ref{def:graph representation} in Section~\ref{sec:graph formulation}, we prepare the reward set $\mc V$ and the cost set $\mathcal E_m, \forall m\in \mc M$ (line $1$).
Since each UAV has different speed $\phi_0^m$, we calculate the edge cost $c_{a(l,t), a(l',t')}^m, \forall ((l,t),(l't')) \in \mc{E}_m$ in \eqref{equ:cost function} for all the UAVs (line $2$).
To obtain the vertex reward ${\varrho} _{k}(z_{k}), \forall k\in \mc K$ in the next step, we need to find the feasible route set $\mathcal S_m$ for each UAV $m\in \mathcal M$ (line $3$), and initialize the routes for all the UAVs $\boldsymbol s$ (line $4$).

\emph{Best response update iteration process:} At the iteration process, each UAV intends to maximize its payoff by choosing an optimal route.
For UAV $m$, we first calculate the number of UAVs $z_k$, who are serving task $k, k\in\mathcal K$ (line $7$).
Based on $z_k, k\in \mc K$, we calculate the vertex reward ${\varrho} _{k}(z_{k}), \forall k\in \mc K$ (line $8$).
Then, we compute the payoff $\widetilde{U}_m(s_m, \boldsymbol s_{-m})$ under current strategy profile $\boldsymbol s$ and update $z_k, k\in \mc K$ (line $9$).
Next, UAV $m$ needs to find the optimal route based on the current strategy of the others. We update the vertex reward ${\varrho} _{k}(z_{k}), \forall k\in \mc K$ and run the SP algorithm to find the optimal route $s_m'$ (line $10$-$11$), so each UAV can compute its best response update within $\mc O(L^3T^2)$ time from Proposition~\ref{PRO:time complexity}.
If the new payoff $\widetilde{U}_m(s_m', \boldsymbol s_{-m})$ has improved (line $13$), the strategy of UAV $m$ will be updated (line $14$).

\emph{Output:} This process repeats until the strategy profile $\bs{s}$ does not change any more, so it is a pure NE.

\section{Performance Evaluations} \label{sec:performance evaluation}
 In this section, we present the numerical results to evaluate the performance of our CRS and DRS schemes.
 We first describe the basic simulation setting in Section~\ref{sec:parameters}.
 Then, we discuss the greedy path (GP) scheme and circular path (CP) scheme in Section~\ref{sec:benchmark scheme}, which serve as the benchmark schemes.
 We compare our proposed DRS scheme with these benchmark schemes in Section~\ref{sec:performance analysis}.

\subsection{Simulation Setting} \label{sec:parameters}

 In our simulations, we consider the network topology as shown in Fig.~\ref{fig:system model graph}. multiple UAVs provide services for $L=9$ seamlessly connected regular hexagonal regions with $R=150$ m side length.
 We consider $T=20$ time slots.
 Moreover, we suppose UAV potential location $\boldsymbol{u}_k$ is the center of each region.
 The UAV control station is in region $3$ (see Fig.~\ref{fig:system model graph}).
 According to PPP, ground users are independent and identically distributed in $L$ regions, and the user transmission probability matrix $p$ in Section~\ref{sec:user demand} has been provided.
 The ground users change their location over duration $T$.
 We consider the rotary-wing UAVs, so that the propulsion energy consumption is given by ~\eqref{equ:power consumption}.
 Note that we assume all the UAVs have the same transmission power $P$.
 For each set of parameters, we run the simulations $65$ times with randomized UAVs' sources and destinations, and user demand in MATLAB.
 Other simulation parameters are listed in Table~\ref{tab:system parameters}.

  \begin{table}[t]
\caption{System Parameters}
\label{tab:system parameters}
\centering
\begin{tabular}{|c|c|c|}
\hline
\footnotesize{\textbf{Parameter}} & \footnotesize{\textbf{Description}} & \footnotesize{\textbf{Value}}\\
\hline
$P$ & \footnotesize{Transmission power of UAV} & $26$ dBm \cite{3GPP}\\
\hline
$H$ & \footnotesize{Altitude of the UAV} & $90$ m \cite{3GPP}\\
\hline
$\theta_0$ & \footnotesize{Antenna beamwidth} & $2.7854$ rad \cite{Backhual,Covercon2}\\
\hline
$f_c$&\footnotesize{Carrier frequency}&$2$ GHz \cite{3GPP}\\
\hline
$N_0$ & \footnotesize{Noise power} & $-96$ dBm \cite{3GPP}\\
\hline
$\alpha$ & \footnotesize{Path-loss exponent} & $2$  \cite{Report}\\
\hline
$\eta_1, \eta_2 $ & \footnotesize{ Path-loss for LOS, NLOS} & $3, 23$ dB \cite{3GPP,Report}\\
\hline
$\psi, \zeta$ & \footnotesize{Environment parameters} & $11.95, 0.14$ \cite{Report}\\
\hline
$\Lambda_0, \Lambda_1, \Lambda_2$ & \footnotesize{Rotary-wing UAV physical property} &$580.65, 790.67, 0.01$ \cite{E2}\\
\hline
$\omega$ & \footnotesize{Tip speed of the rotor blade} & $200$ m/s \cite{E2}\\
\hline
$\chi$ & \footnotesize{\tabincell{c}{Mean rotor induced velocity in hovering (see Equation \\ (2.12) in \cite{37} and Equation (12.1) of \cite{38})}}& $7.2$ \cite{E2}\\
\hline
\end{tabular}
\end{table}

\subsection{Benchmark Schemes} \label{sec:benchmark scheme}

 In our simulation, we compare our DRS scheme with two benchmark schemes.

 \emph{Greedy path (GP) scheme:}
 In the GP scheme, we assume that the UAVs makes a one-hop optimal decision.
 That is, from the current region, each UAV aims to find the next step optimal region, which results in the maximum payoff.
 More specifically, we first assume that the current region and time slot of UAV $m$ is $(l_m,t), l_m\in \mc L, t\in \mc T$.
 We define $(l_m',t'), l_m'\in \mc L, t'\in \mc T$ as a feasible region for UAV $m$ to do the next step decision. Let $\mathcal L'\in \mathcal L$ be the next step feasible region set.
 Next, we calculate the corresponding reward minus cost for each feasible region and find the region $l_m^*\in \mathcal L '$ that results in the maximum payoff.
 Note that for each step UAVs make their decisions one by one.
 Mathematically, the GP scheme can be formulated as
 \begin{equation} \label{equ:gp}
 \begin{array}{rll}
      \displaystyle \arg\max_{l_m^*\in \mc L '} & \displaystyle \rho_{a(l_m,t)}-c_{a(l_m^*,t^*),a(l_m,t)}  & \\
      \text{subject\,to} &  \sigma^m_{a(l_m^*,t^*),a(l_m,t)}-\xi_{a(l_m^*,t^*),a(l_m,t)} \leq 0, \forall m\in\mathcal M.\\
 \end{array}
 \end{equation}
 %


 \emph{Circular path (CP) scheme:}
 In the CP scheme, the UAVs provide services for each region periodically during the duration of $T$.
 For the fairness of the users in different regions, the UAVs have the pre-determined circular flight trajectory \cite{E1,CP}.
 In the simulations, we assume that the UAVs serve the regions in a periodic manner in the pre-determined order of the region.
 For instance, we consider two UAVs providing services for $9$ regions.
 UAV $1$ and UAV $2$ both start from the UAV control station (region $3$), and they also need to go back before time $T$.
 The trajectories for UAV $1$ and UAV $2$ are $3,2,1,4,7,3$ and $3,5,8,9,6,3$, respectively.

\subsection{Performance Analysis} \label{sec:performance analysis}
\begin{figure*}[t]
\hspace{-0.5cm}
\centering
\begin{minipage}[t]{0.4\linewidth}
       \includegraphics[width=7.2cm, trim = 1cm 0.7cm 0cm 0.5cm, clip = true]{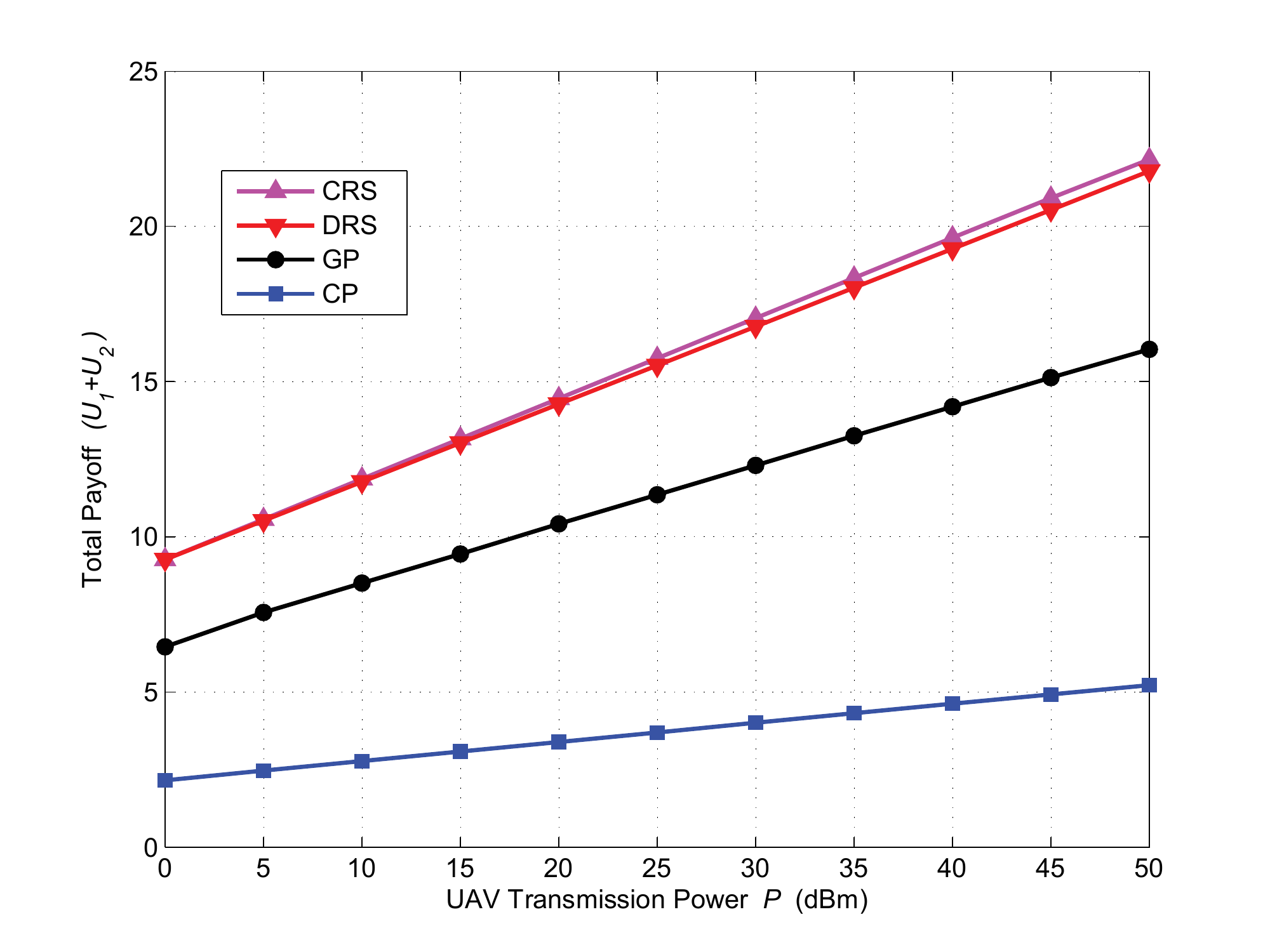}
   \caption{Total payoff for UAV transmission power with $\phi_0^1=\phi_0^2=70$ km/h, $M=2$, and $L=9$.} 
   \label{fig:p_totalpayoff}
\end{minipage}
\quad
\begin{minipage}[t]{0.4\linewidth}
       \includegraphics[width=7.2cm, trim = 1cm 0.7cm 0cm 0.5cm, clip = true]{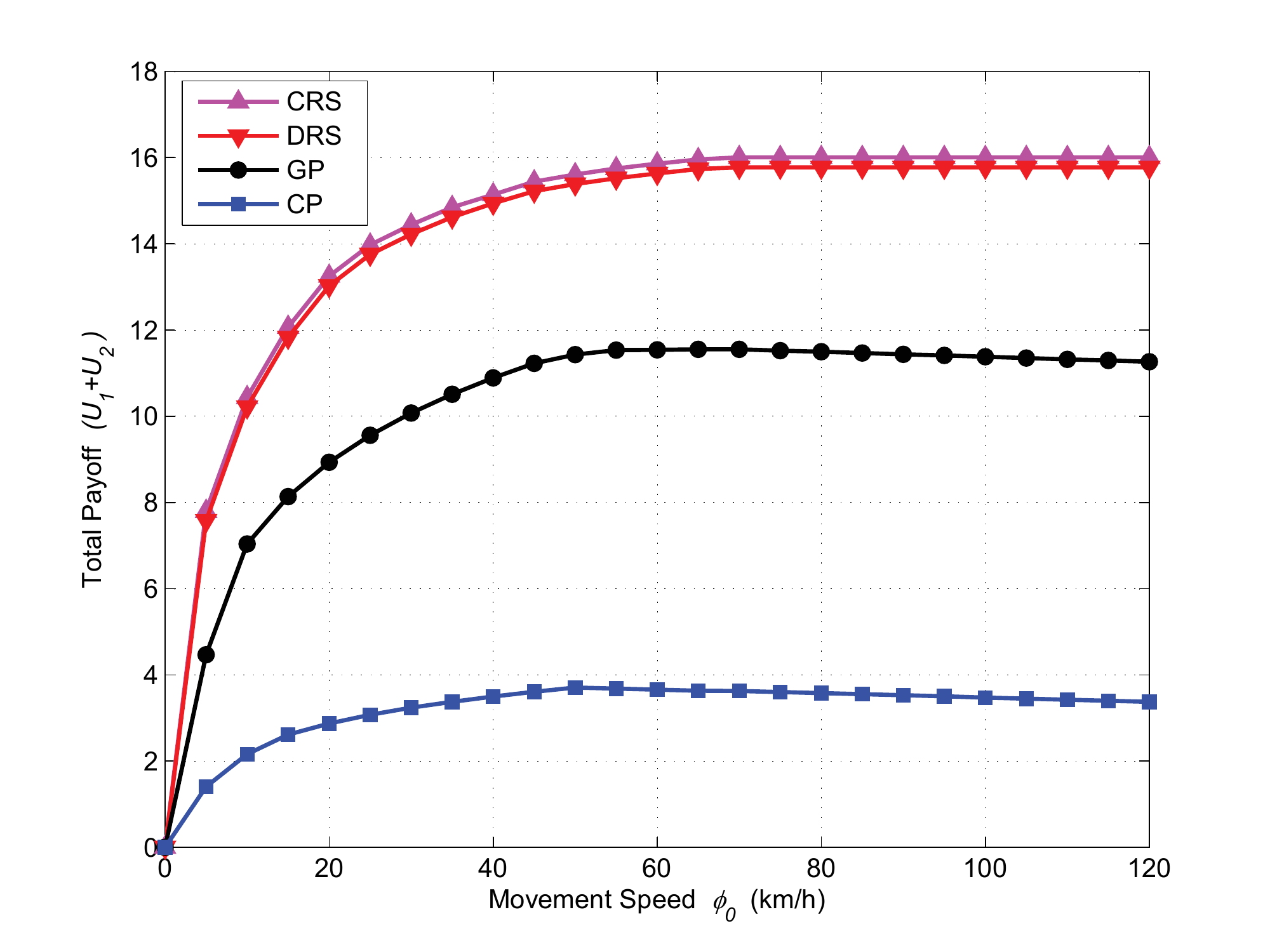}
   \caption{Total payoff for UAV movement speed $\phi_0^1=\phi_0^2$ with $M=2$ and $L=9$.}
   \label{fig:v_totalpayoff}
\end{minipage}
\end{figure*}
\begin{figure*}[ht]
\hspace{-0.5cm}
\centering
\begin{minipage}[t]{0.4\linewidth}
       \includegraphics[width=7.2cm, trim = 1cm 0.7cm 0cm 0.5cm, clip = true]{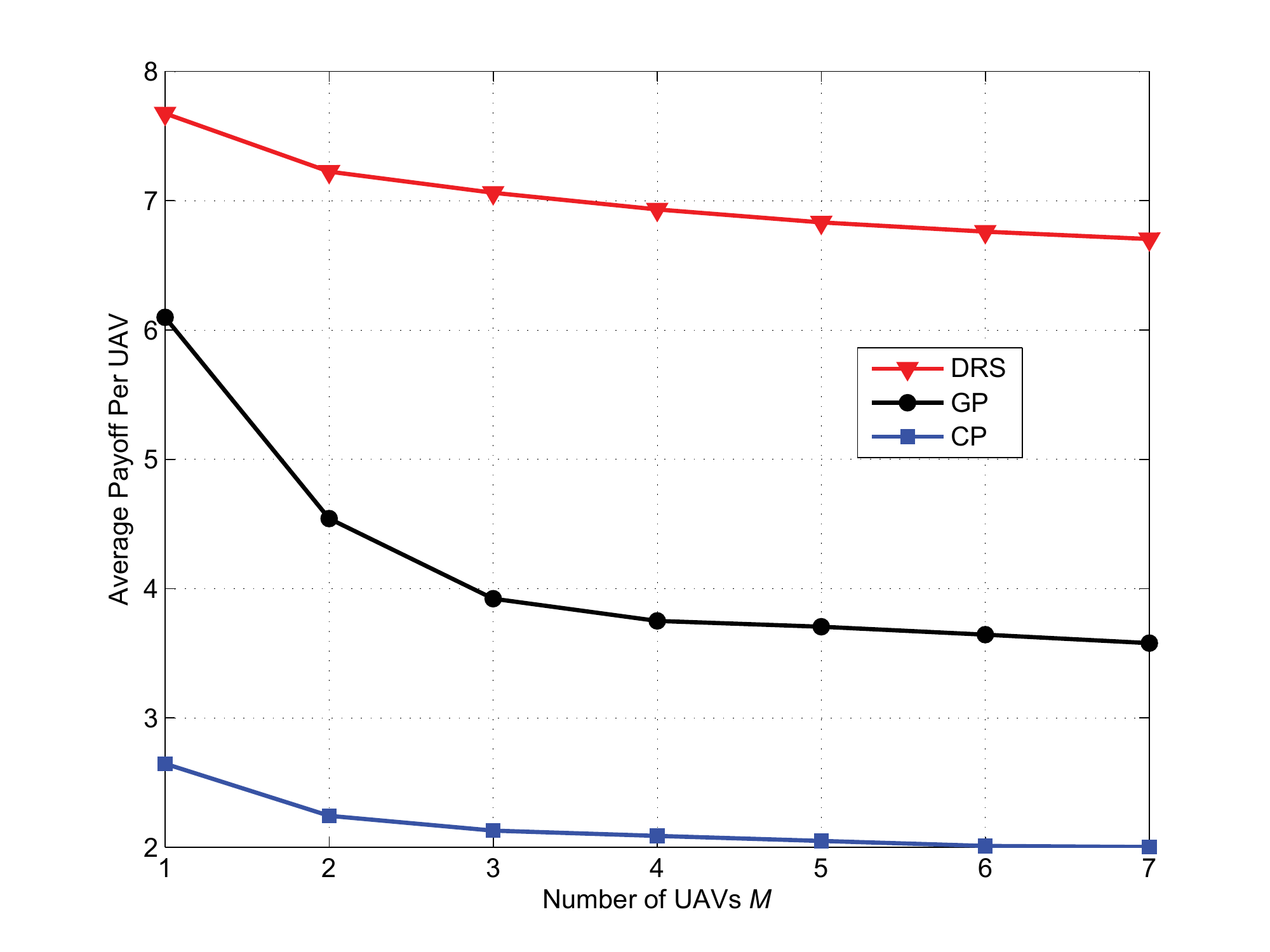}
   \caption{Average payoff per UAV for the number of UAVs $M$ with $\phi_0=50$ km/h and $L=36$.} 
   \label{fig:m_averagepayoff}
\end{minipage}
\quad
\begin{minipage}[t]{0.4\linewidth}
       \includegraphics[width=7.2cm, trim = 1.0cm 0.7cm 0cm 0.5cm, clip = true]{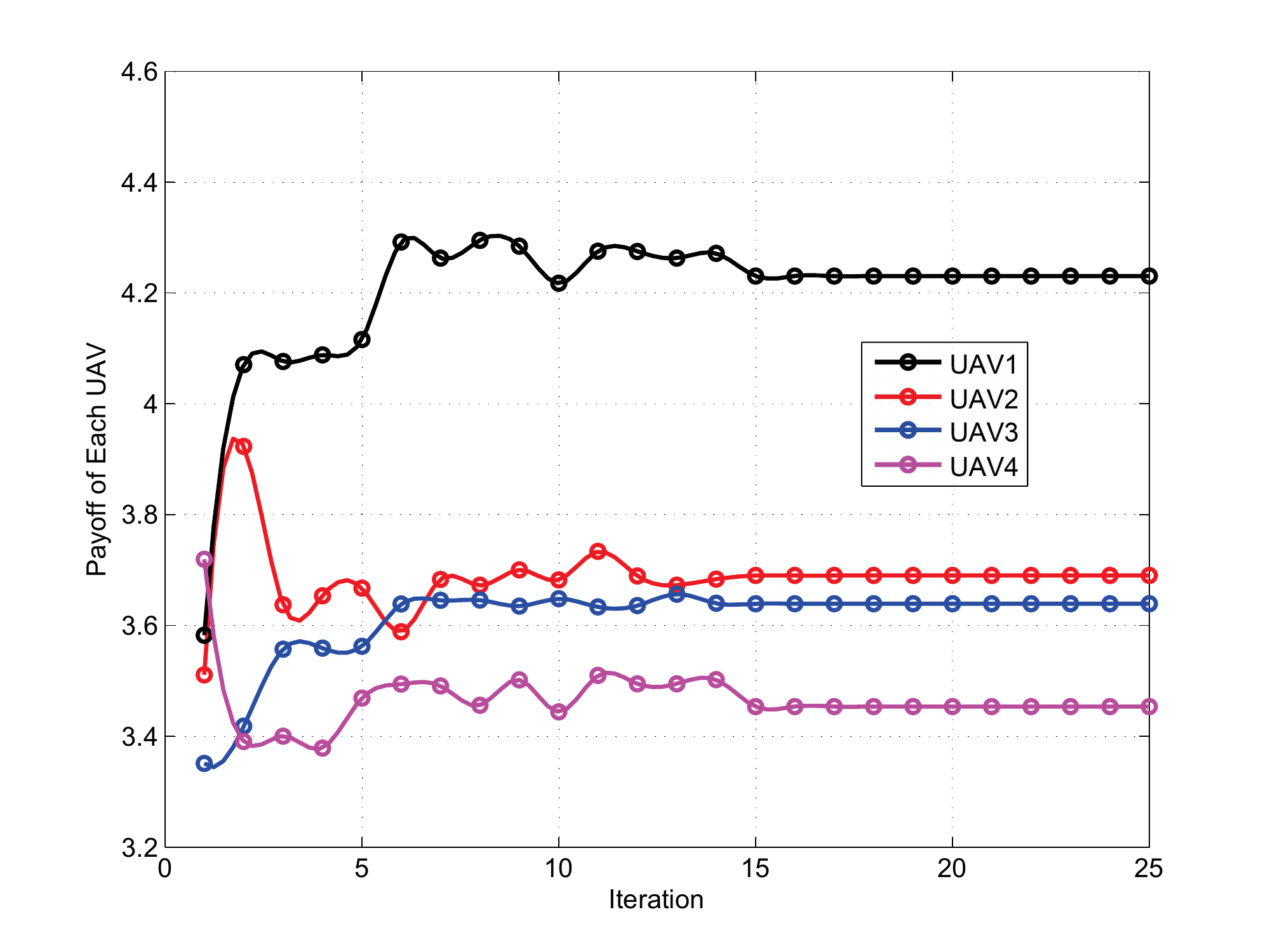}
   \caption{Convergence of DRS scheme for UAV total payoff maximization with $\phi_0^1=\phi_0^2=\phi_0^3=\phi_0^4=50$ km/h, $M=4$, and $L=36$.}
   \label{fig:m_each UAV payoff}
\end{minipage}
\end{figure*}

In this subsection, we provide numerical results to evaluate the performance of our proposed CRS and DRS schemes.

\textbf{Impact of transmission power:}
We study the impact of the UAV transmission power on the total payoff of $M=2$ UAVs.
In Fig.~\ref{fig:p_totalpayoff}, we plot the total payoff against the UAV transmission power $P$ among all four schemes.
First, we observe that the total payoff increases with the transmission power. It is because the average user throughput (i.e., the reward) increases with the transmission power $P$ under the antenna interference mitigation technique   discussed in Section~\ref{sec:channel model}. Moreover,
We can see that our DRS scheme achieves the highest payoff compared to the benchmark schemes. Also, the DRS scheme achieves almost the same payoff as the CRS scheme (optimal scheme) when $P \leq 15$ dBm, and achieves $95\%$ payoff comparing with the CRS scheme when $P = 50$ dBm.
%


\textbf{Impact of movement speed:}
Fig.~\ref{fig:v_totalpayoff} shows the relationship of the total payoff and the UAV movement speed $\phi_0$.
First, we observe that the total payoff increases with the movement speed. It is because the UAVs have more feasible routes as speed $\phi_0$ increases, and more tasks can be served by the UAVs.
However, when the movement speed increases beyond $100$ km/h, the total payoff slightly decreases. It is because the reward (i.e., the average user throughput) will not increase further with a higher speed, while the energy consumption cost $P(\phi_0)$ in \eqref{equ:power consumption} increases with speed $(\phi_0)$.
Overall, we can see that the DRS scheme achieves the highest payoff among the two benchmark schemes.

%

%
\begin{figure}[t]
\centering
       \includegraphics[width=8.3cm, trim = 0cm 0.4cm 0cm 1.0cm, clip = true]{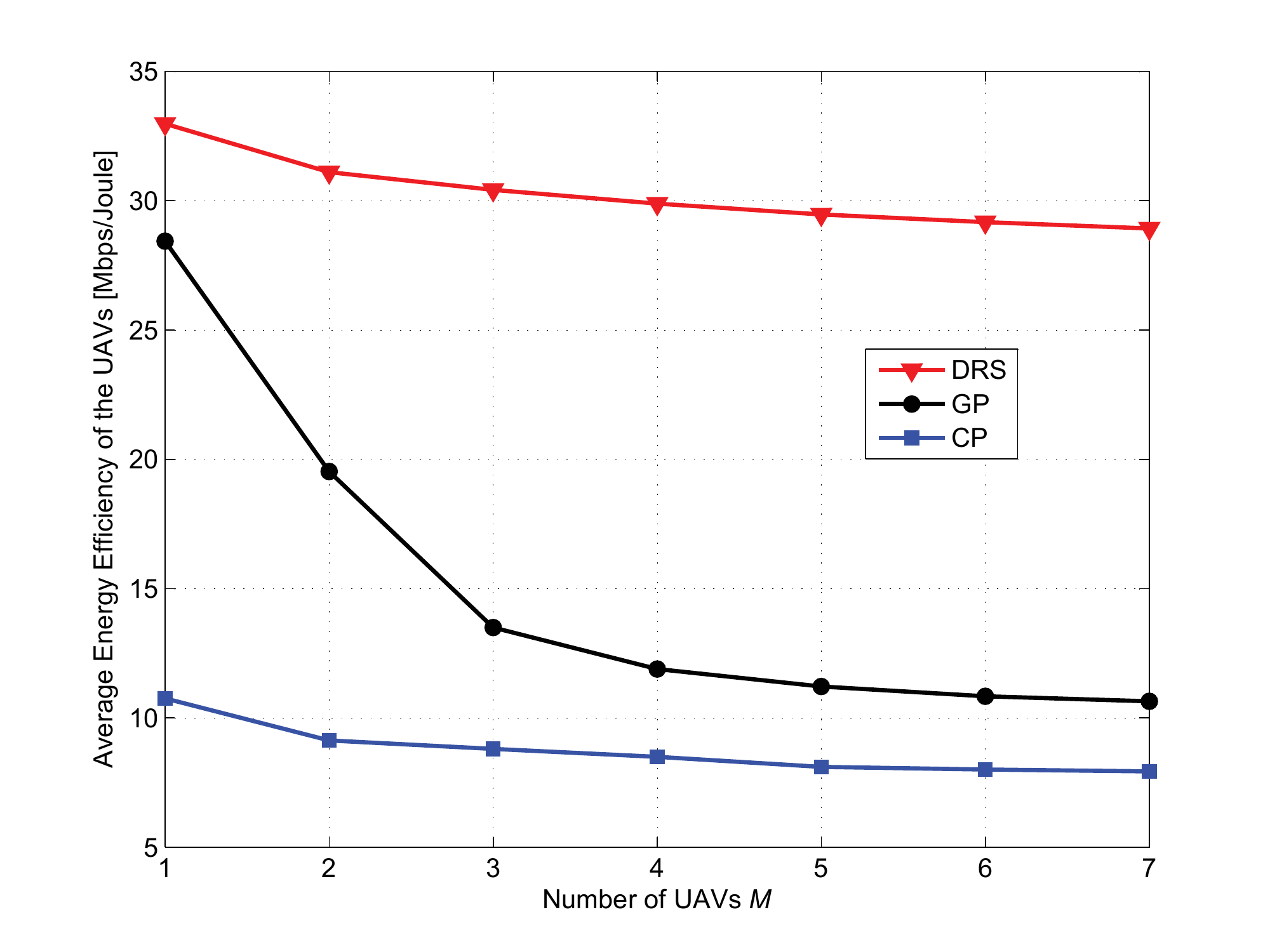}
\caption{Impact of the number of UAVs on the average energy efficiency with $\phi_0=50$ km/h and $L=36$.}
   \label{fig:m_energyefficiency}
\end{figure}
\textbf{Impact of the number of UAVs}\footnote{We do not include the CRS scheme (optimal scheme) in Fig.~\ref{fig:m_averagepayoff} and~\ref{fig:m_energyefficiency}, because it is computationally intensive to consider $M > 3$.}:
Fig.~\ref{fig:m_averagepayoff} shows the impact of the number of UAVs on the average payoff per UAV.
First, we can observe that the average payoff per UAV decreases with the number of UAVs. The reason is due to the increase in interference and contention among the UAVs.
Also, it can be seen that the DRS scheme has the highest average payoff among the GP scheme and the CP scheme.
Since the UAVs do one-hop optimal decisions, they will miss some tasks when making decisions Therefore, the GP scheme has poor performance.

\textbf{Convergence of the DRS scheme:}
In Fig.~\ref{fig:m_each UAV payoff}, we study the convergence of the DRS scheme.
We consider that $M=4$ UAVs providing services for $L=36$ regions with the same speed$\phi_0^1=\phi_0^2=\phi_0^3=\phi_0^4=50$ km/h.
All these UAVs start from the UAV control station (region $3$) at the same time, and they need to come back at time $T$.
We plot the actual payoff $U_m(\bs{s})$ in \eqref{equ:payoff distributed} of each UAV $m$ after each iteration in Algorithm~\ref{Alg:DRS}.
We can see that each UAV's payoff first oscillates and converges finally to an equilibrium point as we have proven in Theorem~\ref{Thm:FIP}.

\textbf{Energy efficiency:}
Fig.~\ref{fig:m_energyefficiency} shows the impact of the number of UAVs on the average energy efficiency, which is defined as the average user throughput divided by the energy consumption.
We can see that the DRS scheme achieves the highest energy efficiency among the GP scheme and the CP scheme.
For the GP scheme, since the UAVs do one-hop optimal decisions, they will miss some tasks when making decisions, which reduces their rewards.
For the CP scheme, the UAVs serve the regions periodically.
The regions with higher rewards have no priority, therefore, the total reward will be reduced, which limits the energy efficiency.

\section{Conclusion} \label{sec:conclusion}
 In this paper, we bring the \emph{user mobility} into the UAVs' trajectory design, while considering the \emph{propulsion energy consumption} and \emph{interference mitigation}.
 Specifically, we defined the average user throughput as the reward for a UAV to serve a region and the propulsion energy consumption as the cost.
 We formulated the UAVs' trajectory design as a route selection problem in an acyclic directed graph.
 For \emph{centralized} trajectory design, we proposed the SP scheme for the single-UAV case and the CRS scheme to systematically compute the \emph{optimal} trajectories for the multiple-UAV case, which is an NP-hard problem.
 Thus, we proposed the DRS scheme for the \emph{distributed} trajectory design and proved that it will converge to an equilibrium point within a finite number of iterations.
 Simulation results showed that the DRS scheme achieves the highest payoff and the energy efficiency as compared with the benchmark GP and CP schemes.

 For the future work, we will consider the impact of the tasks' delay tolerance in some practical scenarios. For example, each user has a buffer to store the tasks have not been executed. It is challenging to solve this problem since the user demand is affected by task generation probability and the delay tolerance time.

\section{Appendix} \label{sec:appendix}
\subsection{Proof of Theorem \ref{Thm:FIP}} \label{sec:A}

 We first need to show that the RSG is a potential game defined as follows:
 \begin{definition}[Potential game] \label{def:potentail function}
 RSG is a potential game if there exists an exact potential function $\Psi(\boldsymbol s)$ such that \cite{Potentialgame}
 \begin{equation}\label{equ:potential function require}
 \begin{aligned}
 \Psi(s_m', \boldsymbol s_{-m})-\Psi(s_m, \bs{s}_{-m})=\widetilde{U}_m(s_m', \boldsymbol s_{-m})-\widetilde{U}_m(s_m, \bs{s}_{-m}), \forall s_m, s_m' \in \mc S_m, m\in \mc M.\\
 \end{aligned}
 \end{equation}
 \end{definition}
  We know that a finite game with an exact potential function has the FIP, which is based on Lemma $2.5$ in \cite{Potentialgame}.

In our RSG, we define the potential function as
\begin{equation} \label{equ:potential function}
    \Psi(\boldsymbol s)= \sum_{(l,t)\in\mathcal V(\bs s)}\sum_{q=1}^{z_{a(l,t)}(\bs s)} {\varrho}_{a(l,t)}(q) - \sum_{m\in \mathcal M}\sum _{((l,t),(l',t'))\in \mathcal E(s_m)}c_{a(l,t),a(l',t')}^m,\\
\end{equation}
where $z_{a(l,t)}(\bs s)$ represents the number of UAVs providing ervices for task $k=a(l,t)$ based on current strategy profile $\bs s$.
We aim to show that this potential function and the payoff function in \eqref{equ:payoff distributed} satisfy \eqref{equ:potential function require}.

To show that $\Psi(\boldsymbol s)$ is a potential function, we first separate the potential function and the payoff function into two parts.
\begin{equation} \label{equ:potential vertex part}
\begin{aligned}
    \Psi^V(\boldsymbol s)&\triangleq\sum_{(l,t)\in\mathcal V(\bs s)}\sum_{q=1}^{z_{a(l,t)}(\bs s)} {\varrho}_{a(l,t)}(q),\\
\end{aligned}
\end{equation}
and
\begin{equation} \label{equ:potential edge part}
\begin{aligned}
    \Psi^E(\boldsymbol s)&\triangleq\sum_{m\in \mathcal M}\sum_{((l,t),(l',t'))\in \mathcal E(s_m)}c_{a(l,t),a(l',t')}^m,\\
\end{aligned}
\end{equation}
so that the potential function in \eqref{equ:potential function} can be represented as
\begin{equation} \label{equ:potential two parts}
\begin{aligned}
    \Psi(\boldsymbol s)=\Psi^V(\boldsymbol s)-\Psi^E(\boldsymbol s).\\
\end{aligned}
\end{equation}
Similarly, we define
\begin{equation} \label{equ:payoff vertex part}
\begin{aligned}
    \widetilde {U}_m^V(\boldsymbol s)\triangleq\sum_{(l,t)\in\mathcal V(s_m)} {\varrho} _{a(l,t)}(z_{a(l,t)}(\bs s)),\\
\end{aligned}
\end{equation}
and
\begin{equation} \label{equ:payoff edge part}
\begin{aligned}
    \widetilde{U}_m^E(\boldsymbol s)\triangleq\sum_{((l,t),(l',t'))\in \mathcal E(s_m)}c_{a(l,t),a(l',t')}^m,\\
\end{aligned}
\end{equation}
such that the payoff function in \eqref{equ:payoff distributed} of UAV $m$ can be represented as
\begin{equation} \label{equ:payoff two parts}
\begin{aligned}
    \widetilde{U}_m(\boldsymbol s)=\widetilde{U}_m^V(\boldsymbol s)-\widetilde{U}_m^E(\boldsymbol s).\\
\end{aligned}
\end{equation}

We know the current strategy profile is $\bs s=(s_1,\dots,s_M)$, and the UAV $m$ changes its strategy from $s_m \in \mc{S}_m$ to $s_m'\in \mathcal S_m$.
Therefore, the new strategy profile is $\bs s'=(s_m',\bs s_{-m})$.
According to~\eqref{equ:potential two parts} and~\eqref{equ:payoff two parts}, we have
\begin{equation} \label{equ:potentail reduce1}
\begin{aligned}
    \Psi(s'_m,\boldsymbol s_{-m})-\Psi(\boldsymbol s)&=\Psi^V(s'_m,\boldsymbol s_{-m})-\Psi^E(s'_m,\boldsymbol s_{-m})-\Psi^V(\boldsymbol s)+\Psi^E(\boldsymbol s)\\
    &=\Psi^V(s'_m,\boldsymbol s_{-m})-\Psi^V(\boldsymbol s)-\left (\Psi^E(s'_m,\boldsymbol s_{-m})-\Psi^E(\boldsymbol s)\right )\\
    &=(\Psi^V(s'_m,\boldsymbol s_{-m})-\Psi^V(\boldsymbol s))-\left (\widetilde{U}_m^E(s'_m,\boldsymbol s_{-m})-\widetilde{U}_m^E(\boldsymbol s)\right ).
\end{aligned}
\end{equation}
The last item $\widetilde{U}_m^E(s'_m,\boldsymbol s_{-m})-\widetilde{U}_m^E(\boldsymbol s)=\Psi^E(s'_m,\boldsymbol s_{-m})-\Psi^E(\boldsymbol s)$ is true, because only the route of UAV $m$ changes.
For the first item, we have
\begin{equation} \label{equ:potentail reduce2}
\begin{aligned}
    \Psi^V(s'_m,\boldsymbol s_{-m})-\Psi^V(\boldsymbol s)&=\sum_{(l,t)\in\mathcal V(\bs s')}\sum_{q=1}^{z_{a(l,t)}(\bs s')} {\varrho}_{a(l,t)}(q)-\sum_{(l,t)\in\mathcal V(\bs s)}\sum_{q=1}^{z_{a(l,t)}(\bs s)} {\varrho}_{a(l,t)}(q)\\
    &=\sum_{(l,t)\in\mathcal V(\bs s')\backslash\mathcal V(s_m)\cup \mathcal V(s'_m)}\sum_{q=1}^{z_{a(l,t)}(\bs s')} {\varrho}_{a(l,t)}(q)-\sum_{(l,t)\in\mathcal V(\bs s)\backslash\mathcal V(s_m)\cup \mathcal V(s'_m)}\sum_{q=1}^{z_{a(l,t)}(\bs s)} {\varrho}_{a(l,t)}(q)\\
    &+\sum_{(l,t)\in\mathcal V(s_m)\cap \mathcal V(s'_m)}\sum_{q=1}^{z_{a(l,t)}(\bs s')} {\varrho}_{a(l,t)}(q)-\sum_{(l,t)\in\mathcal V(s_m)\cap \mathcal V(s'_m)}\sum_{q=1}^{z_{a(l,t)}(\bs s)} {\varrho}_{a(l,t)}(q)\\
    &+\sum_{(l,t)\in\mathcal V(s_m)\backslash \mathcal V(s'_m)}\sum_{q=1}^{z_{a(l,t)}(\bs s')} {\varrho}_{a(l,t)}(q)-\sum_{(l,t)\in\mathcal V(s_m)\backslash \mathcal V(s'_m)}\sum_{q=1}^{z_{a(l,t)}(\bs s)} {\varrho}_{a(l,t)}(q)\\
    &+\sum_{(l,t)\in\mathcal V(s'_m)\backslash \mathcal V(s_m)}\sum_{q=1}^{z_{a(l,t)}(\bs s')} {\varrho}_{a(l,t)}(q)-\sum_{(l,t)\in\mathcal V(s'_m)\backslash \mathcal V(s_m)}\sum_{q=1}^{z_{a(l,t)}(\bs s)} {\varrho}_{a(l,t)}(q)\\
    &=\sum_{(l,t)\in\mathcal V(s'_m)\backslash \mathcal V(s_m)}\sum_{q=1}^{z_{a(l,t)}(\bs s')} {\varrho}_{a(l,t)}(q)-\sum_{(l,t)\in\mathcal V(s_m)\backslash \mathcal V(s'_m)}\sum_{q=1}^{z_{a(l,t)}(\bs s)} {\varrho}_{a(l,t)}(q)\\
    &=\widetilde{U}_m^V(s'_m,\boldsymbol s_{-m})-\widetilde{U}_m^V(\boldsymbol s).
\end{aligned}
\end{equation}
Based on our RSG, $z_{a(l,t)}(\bs s'), (l,t)\in\mathcal V(\bs s')\backslash\mathcal V(s_m)\cup \mathcal V(s'_m)$ is the same as $z_{a(l,t)}(\bs s), (l,t)\in\mathcal V(\bs s)\backslash\mathcal V(s_m)\cup \mathcal V(s'_m)$.
Similarly, $z_{a(l,t)}(\bs s'), (l,t)\in\mathcal V(s_m)\cap \mathcal V(s'_m)$ equals to $z_{a(l,t)}(\bs s), (l,t)\in\mathcal V(s_m)\cap \mathcal V(s'_m)$.
Furthermore, $\mathcal V(s_m) \backslash \mathcal V(s'_m)$ represents the vertices on the route $s_m$ but not on the route $s'_m$. Thus, $z_{a(l,t)}(\bs s'), (l,t)\in\mathcal V(s_m) \backslash \mathcal V(s'_m)$ is $0$, and there is no reward for these vertices.
In addition, $\mathcal V(s'_m) \backslash \mathcal V(s_m)$ represents the vertices on the route $s'_m$ but not on the route $s_m$.
For the same reason, $z_{a(l,t)}(\bs s), (l,t)\in\mathcal V(s_m') \backslash \mathcal V(s_m)$ is $0$.
Therefore, equation~\eqref{equ:potentail reduce2} is established.

Substituting equation~\eqref{equ:potentail reduce2} into equation~\eqref{equ:potentail reduce1}, we have
 \begin{equation}\label{equ:potential function require2}
 \begin{aligned}
 \Psi(s_m', \boldsymbol s_{-m})-\Psi(s_m, \bs{s}_{-m})=\widetilde{U}_m(s_m', \boldsymbol s_{-m})-\widetilde{U}_m(s_m, \bs{s}_{-m}), \forall s_m, s_m' \in \mc S_m, m\in \mc M.\\
 \end{aligned}
 \end{equation}
 \hfill \IEEEQED


\bibliographystyle{IEEEtran}
\bibliography{IEEEabrv,myfile}

\begin{thebibliography}{10}
\providecommand{\url}[1]{#1}
\csname url@samestyle\endcsname
\providecommand{\newblock}{\relax}
\providecommand{\bibinfo}[2]{#2}
\providecommand{\BIBentrySTDinterwordspacing}{\spaceskip=0pt\relax}
\providecommand{\BIBentryALTinterwordstretchfactor}{4}
\providecommand{\BIBentryALTinterwordspacing}{\spaceskip=\fontdimen2\font plus
\BIBentryALTinterwordstretchfactor\fontdimen3\font minus
  \fontdimen4\font\relax}
\providecommand{\BIBforeignlanguage}[2]{{%
\expandafter\ifx\csname l@#1\endcsname\relax
\typeout{** WARNING: IEEEtran.bst: No hyphenation pattern has been}%
\typeout{** loaded for the language `#1'. Using the pattern for}%
\typeout{** the default language instead.}%
\else
\language=\csname l@#1\endcsname
\fi
#2}}
\providecommand{\BIBdecl}{\relax}
\BIBdecl

\bibitem{Tutorial}
M.~{Mozaffari}, W.~{Saad}, M.~{Bennis}, Y.~{Nam}, and M.~{Debbah}, ``A tutorial
  on {UAVs} for wireless networks: {Applications}, challenges, and open
  problems,'' \emph{IEEE Communications Surveys \& Tutorials}, vol.~21, no.~3,
  pp. 2334--2360, third quarter 2019.

\bibitem{Survey1}
A.~{Fotouhi}, H.~{Qiang}, M.~{Ding}, M.~{Hassan}, L.~G. {Giordano},
  A.~{Garcia-Rodriguez}, and J.~{Yuan}, ``Survey on {UAV} cellular
  communications: {Practical} aspects, standardization advancements,
  regulation, and security challenges,'' \emph{IEEE Communications Surveys \&
  Tutorials}, vol.~21, no.~4, pp. 3417--3442, fourth quarter 2019.

\bibitem{Survey2}
L.~{Gupta}, R.~{Jain}, and G.~{Vaszkun}, ``Survey of important issues in {UAV}
  communication networks,'' \emph{IEEE Communications Surveys \& Tutorials},
  vol.~18, no.~2, pp. 1123--1152, second quarter 2016.

\bibitem{Coverage1}
A.~{Al-Hourani}, S.~{Kandeepan}, and S.~{Lardner}, ``Optimal {LAP} altitude for
  maximum coverage,'' \emph{IEEE Wireless Commun. Lett}, vol.~3, no.~6, pp.
  569--572, Dec. 2014.

\bibitem{Coverage2}
M.~{Alzenad}, A.~{El-Keyi}, F.~{Lagum}, and H.~{Yanikomeroglu}, ``3-{D}
  placement of an unmanned aerial vehicle base station {(UAV-BS)} for
  energy-efficient maximal coverage,'' \emph{IEEE Wireless Commun. Lett.},
  vol.~6, no.~4, pp. 434--437, Aug. 2017.

\bibitem{Coverage3}
J.~{Lyu}, Y.~{Zeng}, R.~{Zhang}, and T.~J. {Lim}, ``Placement optimization of
  {UAV-mounted} mobile base stations,'' \emph{IEEE Commun. Lett}, vol.~21,
  no.~3, pp. 604--607, Mar. 2017.

\bibitem{Coverage4}
M.~{Mozaffari}, W.~{Saad}, M.~{Bennis}, and M.~{Debbah}, ``Efficient deployment
  of multiple unmanned aerial vehicles for optimal wireless coverage,''
  \emph{{IEEE} Commun. Lett.}, vol.~20, no.~8, pp. 1647--1650, Aug. 2016.

\bibitem{Loon}
\BIBentryALTinterwordspacing
\emph {Project Loon.} [Online]. Available: \url{https://www.google.com/loon}
\BIBentrySTDinterwordspacing

\bibitem{3GPP}
{3GPP Technical Report 36.777}, ``Technical specification group radio access
  network; {Study} on enhanced {LTE} support for aerial vehicles (release
  15),'' Dec. 2017.

\bibitem{White}
``Unmanned aerial vehicle ({UAV}) utilization of cellular services,'' White
  Paper, ATIS, Aug. 2017.

\bibitem{T1}
Q.~{Wu}, Y.~{Zeng}, and R.~{Zhang}, ``Joint trajectory and communication design
  for {multi-UAV} enabled wireless networks,'' \emph{IEEE Trans. Wireless
  Commun.}, vol.~17, no.~3, pp. 2109--2121, Mar. 2018.

\bibitem{T2}
F.~{Jiang} and A.~L. {Swindlehurst}, ``Optimization of {UAV} heading for the
  ground-to-air uplink,'' \emph{{IEEE} J. Select. Areas Commun.}, vol.~30,
  no.~5, pp. 993--1005, Jun. 2012.

\bibitem{T3}
Y.~{Zeng}, R.~{Zhang}, and T.~J. {Lim}, ``Throughput maximization for
  {UAV-enabled} mobile relaying systems,'' \emph{IEEE Trans. Commun.}, vol.~64,
  no.~12, pp. 4983--4996, Dec. 2016.

\bibitem{T4}
H.~{Wang}, G.~{Ren}, J.~{Chen}, G.~{Ding}, and Y.~{Yang}, ``Unmanned aerial
  vehicle-aided communications: Joint transmit power and trajectory
  optimization,'' \emph{IEEE Wireless Commun. Lett.}, vol.~7, no.~4, pp.
  522--525, Aug. 2018.

\bibitem{T5}
J.~{Lyu}, Y.~{Zeng}, and R.~{Zhang}, ``Cyclical multiple access in {UAV}-aided
  communications: A throughput-delay tradeoff,'' \emph{IEEE Wireless Commun.
  Lett.}, vol.~5, no.~6, pp. 600--603, Dec. 2016.

\bibitem{T6}
Y.~{Zeng}, X.~{Xu}, and R.~{Zhang}, ``Trajectory design for completion time
  minimization in {UAV}-enabled multicasting,'' \emph{IEEE Trans. Wireless
  Commun.}, vol.~17, no.~4, pp. 2233--2246, Apr. 2018.

\bibitem{T7}
J.~{Xu}, Y.~{Zeng}, and R.~{Zhang}, ``{UAV}-enabled wireless power transfer:
  Trajectory design and energy optimization,'' \emph{IEEE Trans. Wireless
  Commun.}, vol.~17, no.~8, pp. 5092--5106, Aug. 2018.

\bibitem{MyGC}
Y.~{Tang}, M.~H. {Cheung}, and T.~M. {Lok}, ``Trajectory design for {UAV}
  assisted wireless networks,'' in \emph{Proc. IEEE GLOBECOM}, 2019.

\bibitem{E1}
Y.~{Zeng} and R.~{Zhang}, ``Energy-efficient {UAV} communication with
  trajectory optimization,'' \emph{{IEEE} Trans. Wireless Commun.}, vol.~16,
  no.~6, pp. 3747--3760, Jun. 2017.

\bibitem{E2}
Y.~{Zeng}, J.~{Xu}, and R.~{Zhang}, ``Energy minimization for wireless
  communication with rotary-wing {UAV},'' \emph{{IEEE} Trans. Wireless
  Commun.}, vol.~18, no.~4, pp. 2329--2345, Apr. 2019.

\bibitem{E3}
C.~{Zhan}, Y.~{Zeng}, and R.~{Zhang}, ``Energy-efficient data collection in
  {UAV} enabled wireless sensor network,'' \emph{IEEE Wireless Commun. Lett.},
  vol.~7, no.~3, pp. 328--331, Jun. 2018.

\bibitem{E4}
D.~{Yang}, Q.~{Wu}, Y.~{Zeng}, and R.~{Zhang}, ``Energy tradeoff in
  ground-to-{UAV} communication via trajectory design,'' \emph{{IEEE} Trans.
  Veh. Technol.}, vol.~67, no.~7, pp. 6721--6726, Jul. 2018.

\bibitem{I1}
U.~{Challita}, W.~{Saad}, and C.~{Bettstetter}, ``Deep reinforcement learning
  for interference-aware path planning of cellular-connected {UAVs},'' in
  \emph{IEEE Int. Conf. on Commun. (ICC)}, 2018.

\bibitem{MMC1}
S.~Gambs, M.-O. Killijian, and M.~N.~n. del Prado~Cortez, ``Next place
  prediction using mobility markov chains,'' in \emph{Proceedings of the First
  Workshop on Measurement, Privacy, and Mobility}, 2012.

\bibitem{MMC2}
------, ``Show me how you move and {I} will tell you who you are,'' in
  \emph{Proceedings of the 3rd ACM SIGSPATIAL International Workshop on
  Security and Privacy in GIS and LBS}, 2010.

\bibitem{CP}
J.~{Lyu}, Y.~{Zeng}, and R.~{Zhang}, ``Spectrum sharing and cyclical multiple
  access in {UAV-aided} cellular offloading,'' in \emph{Proc. IEEE GLOBECOM},
  2017.

\bibitem{Backhual}
\BIBentryALTinterwordspacing
A.~Fotouhi, M.~Ding, and M.~Hassan, ``Dronecells: Improving {5G} spectral
  efficiency using drone-mounted flying base stations,'' \emph{CoRR}, vol.
  abs/1707.02041, 2017. [Online]. Available:
  \url{http://arxiv.org/abs/1707.02041}
\BIBentrySTDinterwordspacing

\bibitem{PPP}
M.~Mozaffari, W.~Saad, M.~Bennis, and M.~Debbah, ``Unmanned aerial vehicle with
  underlaid device-to-device communications: Performance and tradeoffs,''
  \emph{IEEE Trans. Wirel. Commun}, vol.~15, no.~6, pp. 3949--3963, Jun. 2016.

\bibitem{Covercon2}
B.~{Galkin}, J.~{Kibilda}, and L.~A. {DaSilva}, ``Coverage analysis for
  low-altitude {UAV} networks in urban environments,'' in \emph{Proc. IEEE
  {GLOBECOM}}, 2017.

\bibitem{LOSP}
A.~{Al-Hourani}, S.~{Kandeepan}, and A.~{Jamalipour}, ``Modeling air-to-ground
  path loss for low altitude platforms in urban environments,'' in \emph{Proc.
  IEEE {GLOBECOM}}, Austin, TX, USA, 2014.

\bibitem{Chanelgain}
M.~{Mozaffari}, W.~{Saad}, M.~{Bennis}, and M.~{Debbah}, ``Mobile unmanned
  aerial vehicles ({UAVs}) for energy-efficient {Internet of Things}
  communications,'' \emph{IEEE Trans. Wireless Commun.}, vol.~16, no.~11, pp.
  7574--7589, Nov. 2017.

\bibitem{Avergaethput}
Z.~{Wang}, L.~{Duan}, and R.~{Zhang}, ``Adaptive deployment for {UAV}-aided
  communication networks,'' \emph{{IEEE} Trans. Wireless Commun.}, vol.~18,
  no.~9, pp. 4531--4543, Sep. 2019.

\bibitem{Alg}
T.~Cormen, C.~Leiserson, R.~Rivest, and C.~Stein, \emph{Introduction to
  Algorithms}.\hskip 1em plus 0.5em minus 0.4em\relax MIT Press.

\bibitem{Combi}
------, \emph{Algebraic Graph Theory}.\hskip 1em plus 0.5em minus 0.4em\relax
  Cambridge University Press.

\bibitem{Potentialgame}
\BIBentryALTinterwordspacing
D.~Monderer and L.~S. Shapley, ``Potential games,'' \emph{Games and Economic
  Behavior}, vol.~14, no.~1, pp. 124 -- 143, 1996. [Online]. Available:
  \url{http://www.sciencedirect.com/science/article/pii/S0899825696900445}
\BIBentrySTDinterwordspacing

\bibitem{MichaelMBHOC}
M.~H. {Cheung}, R.~{Southwell}, F.~{Hou}, and J.~{Huang}, ``Distributed
  time-sensitive task selection in mobile crowdsensing,'' in \emph{Mobihoc},
  2015.

\bibitem{MichaelArticle}
M.~H. {Cheung}, F.~{Hou}, J.~{Huang}, and R.~{Southwell}, ``Congestion-aware
  {DNS} for integrated cellular and {Wi-Fi} networks,'' \emph{{IEEE} J. Select.
  Areas Commun.}, vol.~35, no.~6, pp. 1269--1281, Jun. 2017.

\bibitem{Report}
{3GPP Technical Report 38.901}, ``Study on channel model for frequencies from
  0.5 to 100 {GHz} (release 14),'' May. 2017.

\bibitem{37}
A.~Bramwell, G.~Done, and D.~Balmford, ``Bramwell's helicopter dynamics,''
  \emph{The Aeronautical Journal (1968)}, vol. 105, no. 1051, p. 534–534, 2001.

\bibitem{38}
A.~Filippone, ``Flight performance of fixed and rotary wing aircraft,''
  \emph{The Aeronautical Journal (1968)}, vol. 111, no. 1123, p. 603–603, 2007.

\end{thebibliography}

\end{document}